\documentclass[11pt]{article}
\usepackage{geometry}                % See geometry.pdf to learn the layout options. There are lots.
\usepackage{graphicx}
\usepackage{amssymb}
\usepackage{mathtools}
\usepackage{epstopdf}
\usepackage{verbatim}
\usepackage{color}
\usepackage{slashed}
\usepackage{bbold}
\usepackage{fullpage}
\usepackage{authblk}
\usepackage[title]{appendix}
\usepackage[colorlinks, citecolor=red, linkcolor=blue]{hyperref}

\newcommand{\Lagr}{\mathcal{L}}
\newcommand{\beq}{\begin{equation}}
\newcommand{\eeq}{\end{equation}}
\newcommand{\bmat}{\begin{pmatrix}}
\newcommand{\emat}{\end{pmatrix}}
\newcommand{\bal}{\begin{align}}
\newcommand{\eal}{\end{align}}

\newcommand{\bm}{\boldsymbol}
\newcommand{\ol}{\overline}
\newcommand{\Order}{\mathcal{O}}
\newcommand{\sg}{\tilde{g}}
\newcommand{\sq}{\tilde{q}}
\newcommand{\snu}{\tilde{\chi}^0}
\newcommand{\sch}{\tilde{\chi}^\pm}
\newcommand{\AY}{\hat{A}}
\newcommand{\ayu}{\hat{A}^u}
\newcommand{\ayd}{\hat{A}^d}
\newcommand{\yu}{Y^u}
\newcommand{\yd}{Y^d}

\DeclareMathOperator{\Tr}{Tr}
\DeclareMathOperator{\cm}{cm}
\DeclareMathOperator{\TeV}{TeV}
\DeclareMathOperator{\GeV}{GeV}
\DeclareMathOperator{\im}{Im}
\DeclareMathOperator{\diag}{diag}

\begin{document}

\title{\begin{flushright}\small{MCTP-14-15} Ê
\end{flushright} \LARGE\bf{Theoretical Prediction and Impact of Fundamental Electric Dipole Moments}}
\author{\normalsize\bf{Sebastian~A.~R.~Ellis\footnote{sarellis@umich.edu}~}}
\author{\normalsize\bf{Gordon~L.~Kane\footnote{gkane@umich.edu}}}
\affil{\it{Michigan Center for Theoretical Physics (MCTP),}\\ \it{Department of Physics, University of Michigan}, \\ \it{Ann Arbor, MI 48109 USA}}
\date{\small{\today}}
\maketitle
\begin{abstract}
\small The predicted Standard Model (SM) electric dipole moments (EDMs) of electrons and quarks are tiny, providing an important window to observe new physics. Theories beyond the SM typically \textit{allow} relatively large EDMs. The EDMs depend on the relative phases of terms in the effective Lagrangian of the extended theory, which are generally unknown. Underlying theories, such as string/M-theories compactified to four dimensions, could predict the phases and thus EDMs in the resulting supersymmetric (SUSY) theory.  Earlier one of us, with collaborators, made such a prediction and found, unexpectedly, that the phases were predicted to be zero at tree level in the theory at the unification or string scale $\sim\Order(10^{16}\GeV)$. Electroweak (EW) scale EDMs still arise via running from the high scale, and depend only on the SM Yukawa couplings that also give the CKM phase. Here we extend the earlier work by studying the dependence of the low scale EDMs on the constrained but not fully known fundamental Yukawa couplings. The dominant contribution is from two loop diagrams and is not sensitive to the choice of Yukawa texture. The electron EDM should not be found to be larger than about $ 5 \times 10^{-30} e\cm$, and the neutron EDM should not be larger than about $5 \times 10^{-29}e\cm$. \ These values are quite a bit smaller than the reported predictions from Split SUSY and typical effective theories, but much larger than the Standard Model prediction. Also, since models with random phases typically give much larger EDMs, it is a significant testable prediction of compactified M-theory that the EDMs should not be above these upper limits. The actual EDMs can be below the limits, so once they are measured they could provide new insight into the fundamental Yukawa couplings of leptons and quarks. We comment also on the role of strong CP violation. EDMs probe fundamental physics near the Planck scale. 
\end{abstract}
\maketitle

\vfill\eject

\tableofcontents

\vfill\eject

%%%%%%%%%%%%%%%%%%%%%%%%%%%%%%%%%%
\section{Introduction}

Strong constraints on CP violation originating from physics beyond the Standard Model (BSM) have been imposed by measurements of electric dipole moments (EDMs) of the electron, neutron and heavy atoms. Thus the implication is that new physics should have generic mechanisms for the suppression of EDMs \cite{Pospelov:2005pr}.

In supersymmetric (SUSY) theories, additional sources of CP violation may arise from complex phases in the soft SUSY breaking parameters \cite{Nilles:1983ge, Haber:1984rc}. If SUSY is only an effective theory for physics at the TeV scale, the phases must be treated as arbitrary, leading to large predictions for EDMs unless the phases are tuned to be small, or there are cancellations. CP violation in SUSY models and the implications for EDM predictions has been studied extensively \cite{Ibrahim:1997gj} - \cite{Ellis:2008zy}. If however, we consider SUSY to be the low energy effective theory of an overarching theory such as a compactified string/M-theory, then there must be some underlying mechanism to predict and relate the various phases.

It is well known that the Electroweak scale CP-violating phase of the Standard Model (SM) cannot provide the source of the CP-violation needed for baryogenesis. The compactified M-theory predicts the required phase also does not arise in the softly broken supersymmetric Lagrangian \cite{Kane:2009kv}. Baryogenesis can arise via the Affleck-Dine mechanism \cite{Affleck:1984fy} at high scales, generated with phases generically present in the super partner's composite flat directions and moduli. The magnitudes of both the baryon number and the dark matter then may arise from moduli decay before nucleosynthesis \cite{Kane:2011ih}. The associated phases are high scale ones that have no effects on EDMs. 

Following on from the results presented in \cite{Kane:2009kv} and the body of work behind it \cite{Acharya:2006} - \cite{Acharya:2010zx}, we now concentrate on analysing the CP violating phases in the effective four-dimensional theory resulting from $\mathcal{N} = 1$ compactifications of M-theory with chiral matter. 

There are two-loop contributions to the EDMs that may be the dominant ones. Even in this case, the phases ultimately arise from the superpotential Yukawas that also give the CKM phase, so we discuss the Yukawa phases first, and turn to the results in Section \ref{Results.SEC}. Readers who want to focus on the upper limits can skip Section \ref{CPVtheory.SEC} on a first reading.

Since the phases in the theory only arise in the Yukawa sector at the high scale, measurements of EDMs also become a useful testing ground for various textures at said scale. We present here an analysis of a variety of different textures and how future measurements may be used to constrain the set of possible choices. 

In Section \ref{ReviewPhases.SEC} we present a review of the results found in \cite{Kane:2009kv} that argue that the dominant CP violating phases are in the superpotential Yukawas. In Section \ref{CPVtheory.SEC} we discuss the sources of CP violation in the theory, as well as present the various textures we investigate and their running. We also show how the phases from the Yukawas enter into the computation of EDMs, summarise the current experimental limits, and discuss the Strong CP contribution. In Section \ref{Results.SEC} we present our results, both for two-loop and one-loop contributions to the EDMs. In Section \ref{Conclusion.SEC} we discuss the upper limits and their interpretation. 

%%%%%%%%%%%%%%%%%CP VIOLATION IN THE CCSSSM%%%%%%%%%%%%%%%%%

\section{Review of Compactified M-Theory prediction of Supersymmetry phases}
\label{ReviewPhases.SEC}
\qquad Here we summarize the arguments from reference \cite{Kane:2009kv} that the high-scale soft-breaking supersymmetry Lagangian from the compactified M-theory leads to the prediction that the dominant CP-violation generating EDMs arises from the phases in the superpotential Yukawas, and thus \textit{has the same source as the CKM phase}. 

In reference \cite{Kane:2009kv} it is shown that terms in the superpotential align with the same phase, leaving just one overall phase, which can be rotated away by a global phase transformation. The K\"ahler potential only depends on the real moduli fields, and the meson condensation $\phi \bar{\phi},$ so it introduces no explicit phases. This is shown in detail in Section IIB of reference \cite{Kane:2009kv}. Basically by removing overall phases one can see that $\partial _{J}K$ and $\partial _{J}\bar{W}$ and therefore $F-$ terms are real. It is also argued in \cite{Kane:2009kv} that although higher order corrections to the K\"ahler potential exist, they do not give rise to new CP-violating phases. This is because in the zero flux sector the superpotential only receives non-perturbative corrections from strong gauge dynamics or membrane instantons. The dynamical alignment of phases still works if these additional terms are subdominant, which is required for the consistency of the moduli stabilization. The hidden sector K\"ahler potential may receive perturbative corrections since there is no non-renormalization theorem for the K\"ahler potential. But the meson field $\phi $ is composed of elementary chiral quark fields $Q$ that are charged under the hidden gauge groups, so higher order corrections \ must be functions of $Q^\dag Q$ by gauge invariance, so such corrections are always functions of $\phi^\dag \phi $ which does not introduce any new phases. The perturbative corrections to the K\"ahler potential are always functions of moduli $z_{i}+\bar{z}_{i}$ which does not introduce any CP violating phases in the soft terms. The dependence on $z_{i} + \bar{z}_{i}$ follows from the shift PQ symmetry of the axion, which is only broken by exponentially suppressed contributions. Thus the result that the CP violating phases in soft parameters are highly suppressed should be quite robust since it only relied on symmetries.

Also, the K\"ahler potential has an approximately flavor diagonal structure because of the presence of U(1) symmetries under which the chiral matter fields are charged. The conical singularities associated with different flavors do not carry the same charges under the U(1)'s in a given basis, which forbids the existence of off-diagonal terms. Such terms can arise when the symmetries are spontaneously broken, but that should be suppressed. Thus the K\"ahler metric is expected to be approximately flavor diagonal at the high scale. As we discuss later, renormalization group running will generate small flavor off-diagonal effects at the EW scale.

Finally, when the superpotential contribution to the overall high scale $\mu$ parameter vanishes, as it does by the Witten mechanism \cite{Witten:2001bf, Acharya:2011te}, the $\mu $ and $B$ parameters are generated by the Giudice-Masiero mechanism \cite{Giudice:1988yz}. \ Then $\mu $ and $B$ have a common phase, but this phase is not physical since it can be eliminated by a U(1)$_{PQ}$ rotation.

Since $\mu$ vanishes if supersymmetry is unbroken and if the moduli are not stabilised $\mu$ is generically of order $\langle\phi\rangle m_{3/2}/M_{Pl}$, typically an order of magnitude suppressed from $m_{3/2}$ \cite{Acharya:2011te}. Including supergravity constraints gives consistency conditions $B=2m_{3/2}$ and $2\mu\tan\beta \approx m_{3/2}$.

\section{CP violation in the Compactified Theory}
\label{CPVtheory.SEC}

Given the results of \cite{Kane:2009kv}, all of the phases in the full Lagrangian originate from the phases of the Yukawa couplings in the underlying superpotential, up to presumably small corrections from the K\"ahler potential. The Yukawa matrices enter the theory through the matter superpotential
\beq
W =  \bar{u}\bm{Y_u} Q H_u - \bar{d} \bm{Y_d} Q H_d - \bar{e} \bm{Y_e} L H_d + \mu H_u H_d
\eeq
where the $\bm{Y_i}$ are $3\times3$ complex matrices in family space. The objects $\bar{u}, Q, \bar{e}, L, H_u$ and $H_d$ are chiral superfields containing the quark, squark, lepton, slepton and Higgs matter fields.  
Then, the contributions to CP violation in the compactified M-theory come entirely from the Yukawa sector of the theory. 

The Yukawa matrices give rise to the quark and lepton masses by the following interaction Lagrangian

\beq
\Lagr_{Yukawa} = Y^u_{ij}\bar{Q}_{Li} H_u u_{Rj} + Y^d_{ij}\bar{Q}_{Li} H_d d_{Rj} + Y^e_{ij}\bar{L}_{Li} H_d e_{Rj} + h.c.
\label{YukawaSM}
\eeq
where $Y^\alpha_{ij},~~\alpha = u,d,e,~~i,j=1,2,3$ are the Yukawa matrices, and $i,j$ are family indices. The matter fields here are SM quarks and leptons. When the Higgs boson gains a vacuum expectation value, the sizes of the eigenvalues of the Yukawa matrices dictate the masses of the quarks or leptons. Diagonalisation of the Yukawa matrices is performed by unitary left-(right-)handed $V_\alpha^{L(R)}$ matrices in flavour space in the Standard Model:
\beq
{V_\alpha^L}^\dagger Y^\alpha V_\alpha^R = Y_\alpha^{diag} \propto \bmat m_{\alpha_1} & 0 & 0 \\ 0 & m_{\alpha_2} & 0 \\ 0 & 0 & m_{\alpha_3} \emat
\eeq
The CKM matrix is defined as $V_{CKM} = {V_u^L}^\dagger V_d^L$, where these are the up and down-type left-handed unitary diagonalisation matrices, so there must be $\Order(1)$ phases in the Yukawa matrices in order to explain the experimentally observed phase of the Cabibbo-Kobayashi-Maskawa (CKM) matrix. When the Yukawa matrices are diagonalised, the phases that were in the original $3 \times 3$ complex matrices are rotated away by the unitary matrices, so that the eigenvalues are real. Therefore, the left-handed unitary matrices which form the CKM matrix carry the phases that were originally in the Yukawas. 

The trilinears that arise in the supersymmetric soft-breaking Lagrangian are defined as $\hat{A}^\alpha_{ij} = A^\alpha_{ij} Y^\alpha_{ij}$, where $A^\alpha_{ij}$ is a general $3\times3$ matrix. Explicitly, the trilinear terms from the soft Lagrangian can then be written as:

\beq
\Lagr_{soft} \sim A^u_{ij}Y^u_{ij}\tilde{\bar{Q}}_{Li} H_u \tilde{u}_{Rj} + A^d_{ij}Y^d_{ij}\tilde{\bar{Q}}_{Li} H_d \tilde{d}_{Rj} + A^e_{ij}Y^e_{ij}\tilde{\bar{L}}_{Li} H_d \tilde{e}_{Rj}
\eeq
where $A^{u,d,e}$ are the trilinear matrices in the gauge eigenstate basis of matter fields, and we are interested in the structure of $Y^\alpha_{ij}$. The matter fields here are the squarks and sleptons.

Rotating to the super-CKM basis is achieved by using the same rotation matrices that diagonalised the Yukawa matrices above, but applying them now to the SUSY squark fields.
\beq
\hat{A}_{SCKM} \equiv {V_\alpha^L}^\dagger A^\alpha Y^\alpha V_\alpha^R
\eeq
where the family indices have been dropped. If the trilinears are not proportional to the Yukawas in the flavour-eigenstate basis, i.e. $A^\alpha_{ij} \not\propto \mathbb{1}_{3\times3}$, the rotation to the super-CKM basis can in itself induce CP-violating phases in the diagonal components of the $\hat{A}$s, giving rise to possible contributions to EDMs. 

We consider the case where the trilinears $\hat{A}$ are not aligned with the Yukawas as a maximally general treatment of CP violation in the theory. 

Additionally, the running of the Yukawas from the high scale to the low scale will mix potential phases in the off-diagonal components into the diagonal elements, thus giving rise to CP-violating phases at the low scale.

%%%%%%%%%%%%%%%%%%%%YUKAWA TEXTURES%%%%%%%%%%%%%%%%%%%

\subsection{Yukawa textures}

The crux of the analysis lies in the determination of viable Yukawa textures. They need to satisfy the requirement that they accurately describe the mass hierarchy exhibited in quarks and leptons, and also that they give the correct CKM angles and phases. 

In order to better understand the structure of a Yukawa texture, we assume that we may decompose it into $\Order(1)$ complex parameters multiplying real parameters giving the relative sizes of the elements of the matrix.

\beq
Y^\alpha_{ij} = \Order(1) \cdot \Lambda_{ij}
\eeq
where $\Lambda_{ij}$ is the matrix of powers in some small parameter $\epsilon \sim \alpha_{GUT}^{1/2} \sim 0.2$, set at the high scale, which will give us the correct hierarchy. While this choice of $\epsilon$ is not the only possible one, it implies a connection between flavour structure and grand unification, and is therefore attractive. The multiplying $\Order(1)$ is a $3\times3$ matrix of magnitude one entries containing the various phases. 

We consider three types of textures in this analysis, symmetric textures with no zeroes, symmetric textures with zeroes, and asymmetric textures with no zeroes, the reason behind these choices being that this allows us to consider a wide variety of possible texture-dependencies. Although this does not cover all possibilities, we find a maximal prediction, so no further EDMs will arise from additional Yukawas. 

Initially we will consider the two possible cases where either the up-type Yukawa matrix, $Y_u$ is diagonal, or the down-type Yukawa matrix, $Y_d$ is diagonal, with the other constrained only by the CKM matrix. We parameterise the CKM matrix in the following manner without loss of generality

\beq
\label{CKMparam.EQ}
V_{CKM}= \Order(1)
\bmat
1 & \epsilon & \epsilon^3
\\
\epsilon & 1 & \epsilon^2
\\
\epsilon^3 & \epsilon^2 & 1
\emat
= \Order(1)
\bmat
1 & 0.2 &0.008
\\
0.2 & 1 & 0.04
\\
0.008 & 0.04 & 1
\emat
\eeq
where the $\Order(1)$ indicates the presence of a matrix of complex parameters of order 1. This is comparable to the experimentally determined values of the CKM matrix
\beq
|V_{CKM}| \sim 
\bmat
0.97 & 0.23 & 0.004 \\
0.23 & 0.97 & 0.04 \\
0.008 & 0.04 & 0.99
\emat
\eeq
We assume the following hierarchy for the quark masses, as seen in \cite{oai:arXiv.org:0811.2417}.

\begin{align}
\label{hierarchy.EQ}
m_u : m_c : m_t ~~&\equiv~~(\epsilon^8 : \epsilon^4 : 1) \times v_u
\\
m_d : m_s : m_b ~~&\equiv~~(\epsilon^5 : \epsilon^3 : 1) \times v_d
\\
m_e : m_\mu : m_\tau ~~&\equiv ~~(\epsilon^8 : \epsilon^4 : 1) \times v_d
\end{align}
where $v_u = \langle H^0_u \rangle$ and $v_d = \langle H^0_d \rangle$ are the vacuum expectation values (VEVs) of the Higgs fields in equation (\ref{YukawaSM}), with the SM VEV defined as $v^2 = v_u^2 + v_d^2$. It should be noted here that we use the up-quark hierarchy for the leptons. While this is non-standard, it is done in order to have a better fit to the experimentally measured masses. 

Below are ratios between the predicted masses and the observed masses for the choice of $\epsilon = 0.26$ for the up-type quarks, $\epsilon = 0.27$ for the down-type quarks and $\epsilon = 0.36$ for the leptons.  Fixing the masses of the top, bottom and $\tau$ to their known values, we find that these choices give
 
\begin{align}
\frac{m_u^{th}}{m_u^{exp}} : \frac{m_c^{th}}{m_c^{exp}} : \frac{m_t^{th}}{m_t^{exp}} ~~&\equiv~~1.21 : 0.61 : 1
\\
\frac{m_d^{th}}{m_d^{exp}} : \frac{m_s^{th}}{m_s^{exp}} : \frac{m_b^{th}}{m_b^{exp}} ~~&\equiv~~1.15 : 0.79 : 1
\\
\frac{m_e^{th}}{m_e^{exp}} : \frac{m_\mu^{th}}{m_\mu^{exp}} : \frac{m_\tau^{th}}{m_\tau^{exp}} ~~&\equiv ~~1.01 : 0.30 : 1
\end{align}
All of these are within order one factors of $\alpha_{GUT}^{1/2}$, so this is compatible with the experimentally measured values for entries in the CKM matrix, given our parameterisation of $V_{CKM}$ in terms of $\epsilon$. The deviation from one can be due to the unknown $\Order(1)$ factors in our decomposition of the Yukawa matrices described above. 

All EDMs will be proportional to some power of $\epsilon$, as will be seen in a later section. Since we are interested in the size of the detectable EDMs, we look for the largest contributions from the Yukawa couplings, which will have the smallest powers of $\epsilon$.

The first  set of textures we consider is derived as shown in appendix A, and is of the form

\beq
Y^u = 
\bmat
\epsilon^8 & \epsilon^5 & \epsilon^3\\
\epsilon^9 & \epsilon^4 & \epsilon^2\\
\epsilon^{11} & \epsilon^6 & 1
\emat
,~~
Y^d =
\bmat
\epsilon^5 & 0 & 0 \\
0 & \epsilon^3 & 0 \\
0 & 0 & 1
\emat
\label{texture1}
\eeq
 where we have taken advantage of being able to perform rotations such that the down-type Yukawa matrix is diagonal. 
 
The same can be repeated where the up-type Yukawa matrix is diagonal, yielding the following textures

\beq
Y^u = 
\bmat
\epsilon^8 & 0 &0\\
0 & \epsilon^4 &0\\
0 & 0 & 1
\emat
,~~
Y^d =
\bmat
\epsilon^5 & \epsilon^4 & \epsilon^3 \\
\epsilon^6 & \epsilon^3 & \epsilon^2 \\
\epsilon^8 & \epsilon^5 & 1
\emat
\label{texture2}
\eeq

The largest predictions for EDMs, arising from the smallest powers in $\epsilon$, will typically arise from terms such as $Y^{(u,d)}_{33}{Y^{(u,d)}_{32}}^\dagger Y^{(u,d)}_{23}$ in the running of the diagonal terms from the high scale, as they involve the $(3,3)$ term which is $1$. This will be seen in section (3.2.1).

The second class of textures we consider are those with zeroes. We study in particular the only five textures with five zeroes found in \cite{Ramond:1993kv} at the high scale, given below

\begin{align}
&Y^u = 
\bmat
0 & \sqrt{2}\epsilon^6 &0\\
\sqrt{2}\epsilon^6 & \epsilon^4 &0\\
0 & 0 & 1
\emat
,~~
Y^d =
\bmat
0 & 2\epsilon^4 & 0 \\
2\epsilon^4 & 2\epsilon^3 & 4\epsilon^3 \\
0 & 4\epsilon^3 & 1
\emat
\label{texture3}
\\
&Y^u = 
\bmat
0 & \epsilon^6 &0\\
\epsilon^6 & 0 &\epsilon^2\\
0 & \epsilon^2 & 1
\emat
,~~
Y^d =
\bmat
0 & 2\epsilon^4 & 0 \\
2\epsilon^4 & 2\epsilon^3 & 2\epsilon^3 \\
0 & 2\epsilon^3 & 1
\emat
\label{texture4}
\\
&Y^u = 
\bmat
0 & 0 &\sqrt{2}\epsilon^4\\
0 & \epsilon^4 &0\\
\sqrt{2}\epsilon^4 & 0 & 1
\emat
,~~
Y^d =
\bmat
0 & 2\epsilon^4 & 0 \\
2\epsilon^4 & 2\epsilon^3 & 4\epsilon^3 \\
0 & 4\epsilon^3 & 1
\emat
\label{texture5}
\\
&Y^u = 
\bmat
0 & \sqrt{2}\epsilon^6 &0\\
\sqrt{2}\epsilon^6 & \sqrt{3}\epsilon^4 &\epsilon^2\\
0 & \epsilon^2 & 1
\emat
,~~
Y^d =
\bmat
0 & 2\epsilon^4 & 0 \\
2\epsilon^4 & 2\epsilon^3 &0 \\
0 & 0 & 1
\emat
\label{texture6}
\\
&Y^u = 
\bmat
0 & 0 &\epsilon^4\\
0 & \sqrt{2}\epsilon^4 &\epsilon^2/\sqrt{2}\\
\epsilon^4 & \epsilon^2/\sqrt{2} & 1
\emat
,~~
Y^d =
\bmat
0 & 2\epsilon^4 & 0 \\
2\epsilon^4 & 2\epsilon^3 &0 \\
0 & 0 & 1
\emat
\label{texture7}
\end{align}
These have a different hierarchy from that of textures 1 and 2, but all of these textures are consistent with the low-energy fermion masses and the CKM matrix elements. Note that since the matrices are symmetric, pairs of zeroes in the off-diagonal components only count as one zero. Since they are all defined at the high scale, we must turn our attention to their running.

\subsection{Running of Yukawa textures}

In this section we consider the running of the Yukawa textures described in the previous section, and look at how the phases from the off-diagonal terms are rotated into the diagonals by said running. Since we start at the high scale with no phases in the diagonal components, the running of the diagonal elements is crucial to understanding the appearance of phases at the low scale.

The evolution of the up and down Yukawa matrices and the trilinear matrices in the MSSM is well known \cite{Martin:1993zk}, and given by,

\begin{align}
\frac{dY^u}{dt} &= \frac{1}{16 \pi^2} Y^u \left\{ 3 \Tr(Y^u {Y^u}^\dagger) + 3 {Y^u}^\dagger Y^u + {Y^d}^\dagger Y^d \right\} + \ldots
\\
\frac{dY^d}{dt} &= \frac{1}{16 \pi^2} Y^d \left\{  \Tr(3Y^u {Y^u}^\dagger + Y^e {Y^e}^\dagger) + 3 {Y^d}^\dagger Y^d + {Y^u}^\dagger Y^u \right\} + \ldots
\\
\frac{d\AY^u}{dt} &= \frac{1}{16 \pi^2} \AY^u \left\{ 3 \Tr(Y^u {Y^u}^\dagger) + 5 {Y^u}^\dagger Y^u + {Y^d}^\dagger Y^d \right\} \\
&+ Y^u \left\{ 6 \Tr(\AY^u {Y^u}^\dagger) + 4 {Y^u}^\dagger \AY^u + 2{Y^d}^\dagger \AY^d \right\}\ + \ldots
\\
\frac{d\AY^d}{dt} &= \frac{1}{16 \pi^2} \AY^d \left\{ \Tr(3Y^d {Y^d}^\dagger + Y^e {Y^e}^\dagger) + 5 {Y^d}^\dagger Y^d + {Y^u}^\dagger Y^u \right\} \\
&+ Y^d \left\{  \Tr(6\AY^d {Y^d}^\dagger + 2 \AY^e {Y^e}^\dagger) + 4 {Y^d}^\dagger \AY^d + 2{Y^u}^\dagger \AY^u \right\}\ + \ldots
\end{align}
where only terms involving the Yukawa matrices are explicitly shown as they are by far the dominant contribution. The trace terms are merely numbers, while the $Y^\dagger Y$ and $Y^\dagger \AY$ terms are matrices that have $\epsilon$ dependencies.

When looking at the evolution of the various terms, we consider the leading terms in $\epsilon$ that come from the off-diagonal terms, as these will be the ones that multiply the phases that are being rotated into the diagonal components. It should be noted that $\epsilon$ is a parameter fixed at the GUT scale that does not run.

\subsubsection{Texture specific running}

In this section, we look at the leading order contributions in $\epsilon$ to the running of the diagonal components of the Yukawa textures from section 2.1 in order to estimate the size of potential phases appearing at the low scale. We consider the running of both the up and down type textures, as each contributes differently.

For the first texture shown in equation (\ref{texture1}), the dominant terms in the running of the $\AY^u_{ii}$ components are:

\begin{align}
&\frac{d\ayu_{11}}{dt} \sim \frac{1}{16 \pi^2} \left[\ayu_{13} \left\{  5 {Y^u}_{33}^\dagger Y^u_{31}\right\} + \yu_{13}\left\{4 {\yu}^{\dagger}_{33}\ayu_{31}\right\} + \ldots \right] \sim \frac{27 A_0 \epsilon^{14}}{16 \pi^2}
\\
&\frac{d\ayd_{11}}{dt} \sim \frac{1}{16 \pi^2} \left[\ayd_{11} \left\{   {Y^u}_{12}^\dagger Y^u_{21}\right\} + \yd_{11}\left\{2 {\yu}^{\dagger}_{12}\ayu_{21}\right\} \right] \sim \frac{3A_0 \epsilon^{23}}{16 \pi^2}
\\
&\frac{d\ayu_{22}}{dt} \sim \frac{1}{16 \pi^2} \left[\ayu_{23} \left\{  5 {Y^u}_{33}^\dagger Y^u_{32}\right\} + \yu_{23}\left\{4 {\yu}^{\dagger}_{33}\ayu_{32}\right\} + \ldots \right] \sim \frac{18 A_0 \epsilon^{8}}{16 \pi^2}
\\
&\frac{d\ayd_{22}}{dt} \sim \frac{1}{16 \pi^2} \left[\ayd_{22} \left\{   {Y^u}_{21}^\dagger Y^u_{12}\right\} + \yd_{22}\left\{2 {\yu}^{\dagger}_{21}\ayu_{12}\right\}  \right] \sim \frac{3A_0 \epsilon^{13}}{16 \pi^2}
\\
&\frac{d\ayu_{33}}{dt} \sim \frac{1}{16 \pi^2} \left[\ayu_{33} \left\{  5 {Y^u}_{32}^\dagger Y^u_{23}\right\} + \yu_{33}\left\{4 {\yu}^{\dagger}_{32}\ayu_{23}\right\} \right] \sim \frac{9 A_0 \epsilon^{4}}{16 \pi^2}
\\
&\frac{d\ayd_{33}}{dt} \sim \frac{1}{16 \pi^2} \left[\ayd_{33} \left\{   {Y^u}_{32}^\dagger Y^u_{23}\right\} + \yd_{33}\left\{2 {\yu}^{\dagger}_{32}\ayu_{23}\right\} \right] \sim \frac{3A_0 \epsilon^{4}}{16 \pi^2}
\label{t1Yu33}
\end{align}

For the texture given in equation (\ref{texture2}), the leading order terms are:

\begin{align}
&\frac{d\ayu_{11}}{dt} \sim \frac{1}{16 \pi^2} \left[\ayu_{11} \left\{ {Y^d}_{12}^\dagger Y^d_{21}\right\} + \yu_{11}\left\{2 {\yd}^{\dagger}_{12}\ayd_{21}\right\} \right] \sim \frac{3 A_0 \epsilon^{20}}{16 \pi^2}
\\
&\frac{d\ayd_{11}}{dt} \sim \frac{1}{16 \pi^2} \left[\ayd_{13} \left\{  5 {Y^d}_{33}^\dagger Y^d_{31}\right\} + \yd_{13}\left\{4 {\yd}^{\dagger}_{33}\ayd_{31}\right\} + \ldots \right] \sim \frac{27A_0 \epsilon^{11}}{16 \pi^2}
\\
&\frac{d\ayu_{22}}{dt} \sim \frac{1}{16 \pi^2} \left[\ayu_{22} \left\{ {Y^d}_{23}^\dagger Y^d_{32}\right\} + \yu_{11}\left\{2 {\yd}^{\dagger}_{23}\ayd_{32}\right\} \right] \sim \frac{3 A_0 \epsilon^{14}}{16 \pi^2}
\\
&\frac{d\ayd_{22}}{dt} \sim \frac{1}{16 \pi^2} \left[\ayd_{23} \left\{  5 {Y^d}_{33}^\dagger Y^d_{32}\right\} + \yd_{23}\left\{4 {\yd}^{\dagger}_{33}\ayd_{32}\right\} + \ldots \right] \sim \frac{18A_0 \epsilon^{7}}{16 \pi^2}
\\
&\frac{d\ayu_{33}}{dt} \sim \frac{1}{16 \pi^2} \left[\ayu_{33} \left\{ {Y^d}_{32}^\dagger Y^d_{23}\right\} + \yu_{33}\left\{2 {\yd}^{\dagger}_{32}\ayd_{23}\right\} \right] \sim \frac{3 A_0 \epsilon^{4}}{16 \pi^2}
\\
&\frac{d\ayd_{33}}{dt} \sim \frac{1}{16 \pi^2} \left[\ayd_{33} \left\{  5 {Y^d}_{32}^\dagger Y^d_{23}\right\} + \yd_{33}\left\{4 {\yd}^{\dagger}_{32}\ayd_{23}\right\} + \right] \sim \frac{9A_0 \epsilon^{4}}{16 \pi^2}
\label{t2Yd33}
\end{align} 

As mentioned earlier, the leading order terms, here of  $\Order(\epsilon^4)$ for both the textures in equations (\ref{texture1}) and (\ref{texture2}), arise from the terms of the form $Y^{(u,d)}_{33}{Y^{(u,d)}_{32}}^\dagger Y^{(u,d)}_{23}$. Note that only the leading results for the remaining textures are presented here, and the explicit matrix elements that enter in the running of the other textures are not shown. 

For the texture given in equation (\ref{texture3}), the lowest order terms in $\epsilon$ are 
\begin{align}
&\frac{d\ayu_{11}}{dt} \sim \frac{1}{16 \pi^2} ( 12\sqrt{2} A_0 \epsilon^{13} )
\\
&\frac{d\ayd_{11}}{dt} \sim \frac{1}{16 \pi^2} ( 72 A_0 \epsilon^{11} )
\\
&\frac{d\ayu_{22}}{dt} \sim \frac{1}{16 \pi^2} ( 60 A_0 \epsilon^{10} )
\\
&\frac{d\ayd_{22}}{dt} \sim \frac{1}{16 \pi^2} ( 144 A_0 \epsilon^6 )
\\
&\frac{d\ayu_{33}}{dt} \sim \frac{1}{16 \pi^2} ( 48 A_0 \epsilon^6 )
\\
&\frac{d\ayd_{33}}{dt} \sim \frac{1}{16 \pi^2} ( 288 A_0 \epsilon^6 )
\end{align}

For the texture given in equation (\ref{texture4}), the lowest order terms are
\begin{align}
&\frac{d\ayu_{11}}{dt} \sim \frac{1}{16 \pi^2} ( 12 A_0 \epsilon^{13} )
\\
&\frac{d\ayd_{11}}{dt} \sim \frac{1}{16 \pi^2} ( 72 A_0 \epsilon^{11} )
\\
&\frac{d\ayu_{22}}{dt} \sim \frac{1}{16 \pi^2} ( 9 A_0 \epsilon^{4} )
\\
&\frac{d\ayd_{22}}{dt} \sim \frac{1}{16 \pi^2} ( 6 A_0 \epsilon^5 )
\\
&\frac{d\ayu_{33}}{dt} \sim \frac{1}{16 \pi^2} ( 18 A_0 \epsilon^4 )
\\
&\frac{d\ayd_{33}}{dt} \sim \frac{1}{16 \pi^2} ( 3 A_0 \epsilon^4 )
\label{t4Yu33a}
\end{align}

For the texture in equation (\ref{texture5}), the lowest order terms are
\begin{align}
&\frac{d\ayu_{11}}{dt} \sim \frac{1}{16 \pi^2} ( 18 A_0 \epsilon^{8} )
\\
&\frac{d\ayd_{11}}{dt} \sim \frac{1}{16 \pi^2} ( 72 A_0 \epsilon^{11} )
\\
&\frac{d\ayu_{22}}{dt} \sim \frac{1}{16 \pi^2} ( 60 A_0 \epsilon^{10} )
\\
&\frac{d\ayd_{22}}{dt} \sim \frac{1}{16 \pi^2} ( 144 A_0 \epsilon^6 )
\\
&\frac{d\ayu_{33}}{dt} \sim \frac{1}{16 \pi^2} ( 48 A_0 \epsilon^6 )
\\
&\frac{d\ayd_{33}}{dt} \sim \frac{1}{16 \pi^2} ( 288 A_0 \epsilon^6 )
\end{align}

For the texture in equation (\ref{texture6}), the lowest order terms are
\begin{align}
&\frac{d\ayu_{11}}{dt} \sim \frac{1}{16 \pi^2} ( 24/\sqrt{2} A_0 \epsilon^{13} )
\\
&\frac{d\ayd_{11}}{dt} \sim \frac{1}{16 \pi^2} ( 72 A_0 \epsilon^{11} )
\\
&\frac{d\ayu_{22}}{dt} \sim \frac{1}{16 \pi^2} ( 9 A_0 \epsilon^{4} )
\\
&\frac{d\ayd_{22}}{dt} \sim \frac{1}{16 \pi^2} ( 6 A_0 \epsilon^7 )
\\
&\frac{d\ayu_{33}}{dt} \sim \frac{1}{16 \pi^2} ( 18 A_0 \epsilon^4 )
\\
&\frac{d\ayd_{33}}{dt} \sim \frac{1}{16 \pi^2} ( 3 A_0 \epsilon^4 )
\end{align}

For the texture in equation (\ref{texture7}), the lowest order terms are
\begin{align}
&\frac{d\ayu_{11}}{dt} \sim \frac{1}{16 \pi^2} ( 9 A_0 \epsilon^{8} )
\\
&\frac{d\ayd_{11}}{dt} \sim \frac{1}{16 \pi^2} ( 3\sqrt{2} A_0 \epsilon^{10} )
\\
&\frac{d\ayu_{22}}{dt} \sim \frac{1}{16 \pi^2} ( 9/2 A_0 \epsilon^{4} )
\\
&\frac{d\ayd_{22}}{dt} \sim \frac{1}{16 \pi^2} ( 3 A_0 \epsilon^7 )
\\
&\frac{d\ayu_{33}}{dt} \sim \frac{1}{16 \pi^2} ( 9 A_0 \epsilon^4 )
\\
&\frac{d\ayd_{33}}{dt} \sim \frac{1}{16 \pi^2} ( 3/2 A_0 \epsilon^4 )
\end{align}

%%%%%%%%%%%%%%%%%%%%EDMS FROM YUKAWAS%%%%%%%%%%%%%%%%%
\subsection{Translating from Yukawas to EDMs}
\label{operators.SEC}
Having described various Yukawa textures and their running, we now concentrate on demonstrating how the phases will enter into a computation of the EDMs. 

In the MSSM, the important CP-odd terms in the Lagrangian are

\beq
\begin{split}
\Lagr_{CP-odd} \supset &-\sum_{q=u,d,s} m_q \ol{q}(1+i\theta_q \gamma_5)q + \theta_G \frac{\alpha_s}{8\pi}G \tilde{G}
\\
& -\frac{i}{2} \sum_{q=u,d,s} (d_q^E \ol{q}F^{\mu\nu}\sigma_{\mu\nu}\gamma_5 q + d^C_q \ol{q}g_s T^a G^{a\mu\nu} \sigma_{\mu\nu} \gamma_5 q)
\\
&-\frac{1}{6}d^G_q f_{abc} G_{a\mu\rho} G^\rho_{b\nu} G_{c\lambda\sigma} \epsilon^{\mu\nu\lambda\sigma}
\end{split}
\label{EDMLagr}
\eeq
where $\theta_G$ is the QCD $\theta$ angle, the second line contains dimension five operators, generated by CP violation in the SUSY breaking sector and evolved down to $\sim 1$ GeV. The coefficients $d_q^{E,C}$ correspond to the quark electric and chromo-electric dipole moments (EDM, CEDM) respectively. The last line contains the gluonic dimension six Weinberg operator, to which all other purely gluonic $P$- or $T$-odd operators are proportional \cite{Weinberg:1989dx}. 

The explicit expressions for the SUSY contributions to EDMs are given in Appendix B. The phases appear only in the tri-linear $\hat{A}$ parameters in our theory, and after RG evolution and the rotation to the super-CKM basis, they then appear in the off-diagonal elements of the squark mass matrices.

\beq
\delta(m^2_{\sq})_{ii}^{LR} = v_q((\hat{A}^q_{SCKM})_{ii} - \mu^* Y_{ii}^qR_q)
\eeq
with $R_q = \cot\beta , ~(\tan\beta)$ for $I_3 = 1/2, ~(-1/2)$ as in Appendix B, and $v_{u(d)} = v\sin\beta (v \cos \beta)$. 

The off-diagonal elements of the squark mass matrices enter the expressions for the SUSY EDM contributions as shown in detail in Appendix B. Thus, we find that the EDM contribution  $d_i \propto \im(\hat{A}_{SCKM})$ depends on the phases in the diagonal terms in the trilinears. 

%%%%%%%%%%%%%%%%%%%%EDMS AND CURRENT LIMITS%%%%%%%%%%%%%%%

\subsection{Electric Dipole Moments and Current Experimental Limits}

We now summarize the experimental results on the electron, neutron and mercury EDMs. In minimal SUSY models, the electron EDM arises from one-loop diagrams with chargino and neutralino exchange, as well as two-loop contributions. Hence we can make the decomposition
\beq
d_e^E = d_e^{\chi^\pm} + d_e^{\chi^0} + d_e^{2L}
\eeq
%noting that what is actually measured is the atomic EDM of Thallium $d_{Tl}$, from which the electron EDM is derived. 
%
%\beq
%d_{Tl} = -585 \times d_e^E - 8.5 \times 10^{-19} e \cm (C_S \TeV^2)
%\eeq
%where $C_S$ is the coefficient of the effective operator $\bar{e}i\gamma_5e\bar{N}N$. The current bound on $|d_{Tl}| < 9 \times 10^{-25}e\cm$ implies an upper limit on the electron EDM
The current experimental upper bound on the electron EDM is \cite{Baron:2013eja}
\beq
|d_e^E| < 8.7 \times 10^{-29}e \cm
\eeq

Calculating the neutron EDM requires assumptions about the internal structure, such that there are two possible approaches, the chiral model, and the parton model approach. We will restrict ourselves to the chiral model approach, although a combination could be done in a future study. The neutron EDM can be decomposed by use of the SU(6) coefficients into 

\beq
d_n = \frac{4}{3}d_d - \frac{1}{3}d_u
\label{nedm}
\eeq
which then requires estimation of the quark EDMs, which can be achieved via a naive dimensional analysis, such that
\beq
d_q = \eta^E d_q^E + \eta^C\frac{e}{4\pi} d^C_q + \eta^G \frac{e\Lambda}{4\pi}d^G
\eeq
where $\Lambda \sim 1.19$ GeV is the chiral symmetry breaking scale, and the coefficients are the QCD correction factors, given by $\eta^E = 1.53$, $\eta^C \sim \eta^G \sim 3.4$, as found in \cite{Ibrahim:1997gj}. The contributions from SUSY come from 1-loop gluino, chargino and neutralino exchange, as well as 2-loop contributions, leading to the decomposition
\beq
d_q^{E,C} = d_q^{\tilde{g}(E,C)} + d_q^{\chi^\pm(E,C)} + d_q^{\chi^0(E,C)} + d_q^{2L}
\eeq
and two-loop gluino quark squark diagrams which generate $d_q^G$. The current experimental limit on the neutron EDM is \cite{Baker:2006ts}
\beq
|d_n| < 3 \times 10^{-26}e\cm
\eeq

The mercury EDM results mostly from T-odd nuclear forces in the mercury nucleus, which induce an interaction of the type $(\bm{I}\cdot \nabla)\delta(\bm{r})$ between the electron and the nucleus of spin $\bm{I}$. The T-odd forces themselves arise due to the effective four-fermion interaction $\bar{p}p\bar{n}i\gamma_5n$ \cite{Abel:2001vy}. The current theoretical estimate is given by

\beq
d_{Hg} = -7.0 \times 10^{-3}e(d_d^C -d_u^C -0.012d_s^C) + 10^{-2} \times d_e
\eeq
where the contribution from the strange quark CEDM is included. The experimental bound currently stands at \cite{Griffith:2009zz}

\beq
|d_{Hg}| < 3.1 \times 10^{-29} ~e\cm
\eeq

\subsubsection{Strong CP contribution}

A possible source of hadronic EDMs in the Standard Model comes from the $\theta-$term of QCD. This contribution is shown in equation (\ref{EDMLagr}). The limits on the EDMs of the neutron and Mercury can be expressed in terms of this $\theta$ parameter as follows
\begin{align}
\label{EDMtheta}
d_n &\sim 3 \times 10^{-16}\theta ~e \cm
\nonumber
\\
|d_{Hg}| &\sim \Order(10^{-18}-10^{-19})\theta ~e \cm
\end{align}

The contribution to the electron EDM, on the other hand, comes from electroweak interactions. Thus, with measurements of the neutron EDM and electron EDM, the strong and weak contributions can hopefully be separated, and $\theta$ can also be measured. Our upper limit on the electron EDM is not affected by the strong CP violation, but for the neutron or Mercury,  there could be a strong CP contribution that increases the EW contribution above the EW upper limit. With sufficient data it may be possible to untangle these.  

In the SM, a first analysis of the renormalisation of the $\theta$ parameter \cite{Ellis:1978hq} found that the first renormalisation occurs only at $\Order(\alpha^2)$, which would give a value of $\theta \sim \Order(10^{-16})$. A subsequent detailed analysis  yielded a smaller value of $\theta \sim \Order(10^{-19})$ \cite{Khriplovich:1985jr}. In this case, from equation (\ref{EDMtheta}), we can see that the strong contribution to the neutron EDM from $\theta$ renormalisation would be $\Order(10^{-32}-10^{-35})$.

An estimate of electroweak renormalisation contributions to $\theta$ in SUSY is presented in \cite{Ellis:1982tk}, where it is discussed that $\theta$ is expected to be small, given that relevant phases are small, and $M_{\sq} \gg \Order(100 \GeV)$. Thus, observation of a neutron or Mercury EDM should likely be interpreted as the Electroweak one we estimate in this paper, but needs detailed confirmation.

Solutions of the Strong CP problem in string theory have been studied for example in \cite{Acharya:2010zx, Bobkov:2010rf}. In this case a combination of the imaginary parts of the moduli fields is the QCD axion and solves the Strong CP problem. However, in the presence of non-perturbative contributions, the minimum of the axion potential need not be zero, and $\theta$ can have both strong and electroweak contributions \cite{Acharya:2010zx,Svrcek:2006yi}.

%%%%%%%%%%%%%%%%%%%%%%%%RESULTS%%%%%%%%%%%%%%%%%%%%

\section{Results}
\label{Results.SEC}

Within the framework we are considering of compactified M-Theory, the general structure of SUSY breaking parameters is as follows. The gravitino mass is essentially $\sim F_\phi / M_{Pl}$, which puts it naturally in the range of 25-100 TeV \cite{Acharya:2007}. The $F$-terms of the moduli are suppressed with respect to $F_\phi$, and since the gauge kinetic function for the visible sector depend only on the moduli, it is easy to check that the gaugino masses are suppressed relative to the gravitino mass. Scalar masses, on the other hand, are not suppressed relative to the gravitino mass unless the visible sector is sequestered from the SUSY breaking sector, which is not generic in M-Theory \cite{Acharya:2008}. Thus the scalar masses and trilinears turn out to be of $\Order(M_{3/2}) \gtrsim \Order(50)$ TeV. Due to the K\"ahler metric being approximately diagonal in the flavor indices, the scalar mass matrix is roughly diagonal, with suppressed off-diagonal components. In the following, we consider electroweakinos with masses $\lesssim 600$ GeV, scalar masses $\sim 50$ TeV, with $B$ and trilinear parameters of the same order. The $\mu$ parameter is expected to be suppressed compared to the gravitino mass by an order of magnitude \cite{Acharya:2011te}.

We estimate the contribution to the EDMs of the electron, neutron and mercury from the operators described in Section \ref{operators.SEC} in our chosen M-theory framework. 

\subsection{Dominant two-loop contributions}

\begin{figure}[h!]
\centering
\includegraphics[scale=0.3]{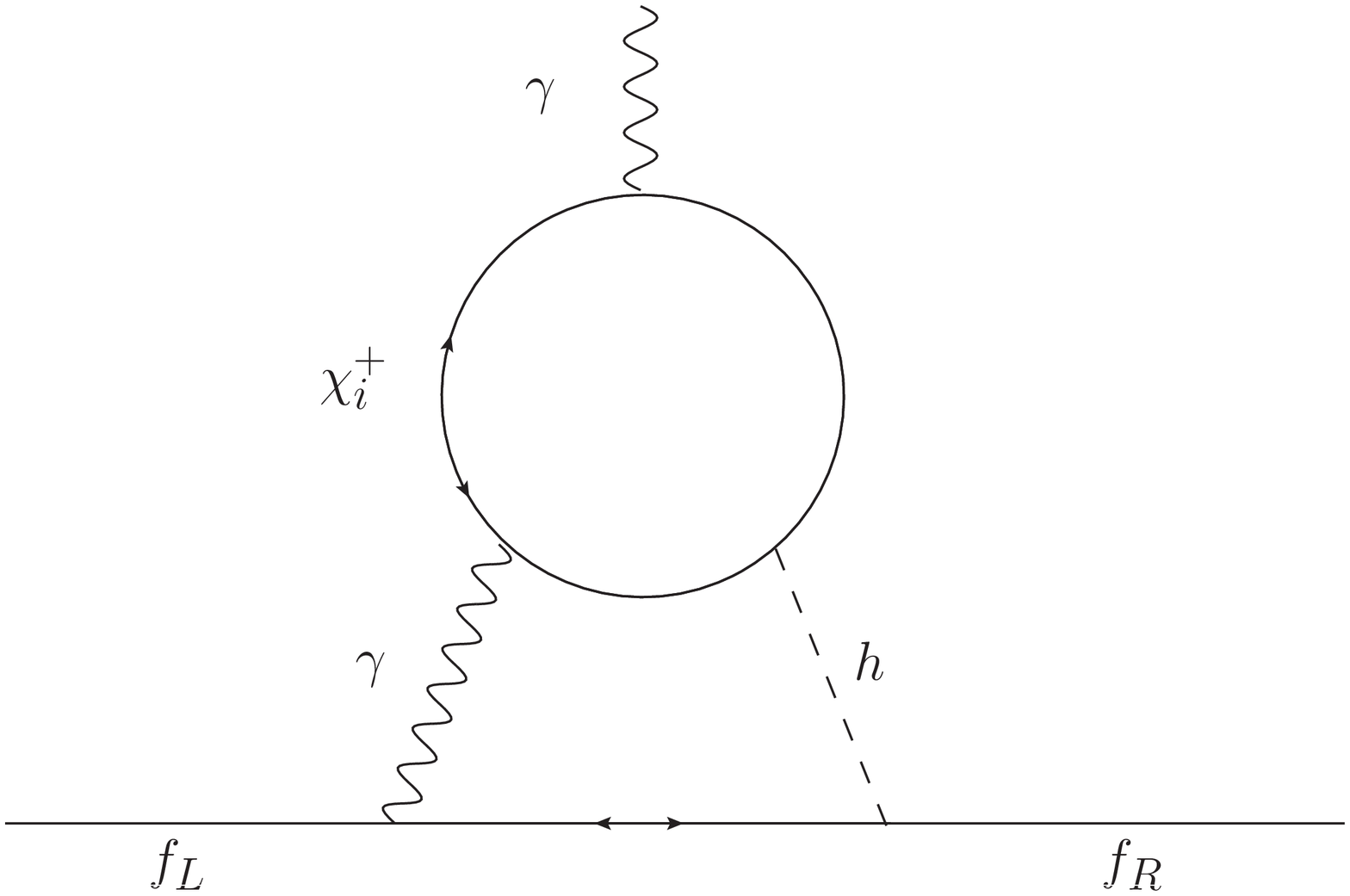}
\caption{An example of a two-loop graph which contributes to fermion EDMs, with charginos running in the inner loop, $\gamma$ and higgs in the outer loop.}
\label{GammaHiggs.FIG}
\end{figure}

There exist two-loop diagrams which could give large contributions to EDMs in supersymmetric models \cite{Chang:1998uc}, \cite{BowserChao:1997bb}-\cite{Li:2008kz}. For example, the diagrams considered in \cite{ArkaniHamed:2004yi, Giudice:2005rz}, one of which is shown in Fig. \ref{GammaHiggs.FIG}, could potentially give large EDMs, as they are not suppressed by the heavy scalar masses, but rather depend on the charginos and neutralinos running in the loops. 

Their contribution to the fermion EDM would be given by 

\beq
d_f = d_f^{\gamma H} + d_f^{ZH} + d_f^{WW}
\eeq
where
\begin{align}
\label{dGR.EQ}
\nonumber d_f^{\gamma H} &= \frac{eQ_f \alpha^2}{4\sqrt{2} \pi^2 s^2_W} \im(D_{ii}^R) \frac{m_f M_i^+}{M_W m_H^2} f_{\gamma H}(r^+_{i H})
\\
\nonumber d_f^{ZH} &= \frac{e(T_{3f_L} -2s^2_WQ_f) \alpha^2}{16\sqrt{2} \pi^2 c^2_W s^4_W} \im(D_{ij}^R G_{ji}^R - D_{ij}^L G_{ji}^L) \frac{m_f M_i^+}{M_W m_H^2} f_{Z H}(r_{ZH},r^+_{i H},r^+_{j H})
\\
d_f^{WW} &=\frac{eT_{3f_L} \alpha^2}{8 \pi^2 s^4_W} \im(C_{ij}^L C_{ij}^{R*})  \frac{m_f M_i^+ M_j^0}{M_W^4} f_{WW}(r^+_{iW}, r^0_{jW})
\end{align}
where
\begin{align}
\label{DiagMatrices.EQ}
\nonumber G^L_{ij} &= V_{iW^+}c_{W^+} V^\dagger_{W^+ j} + V_{i h_u^+} c_{h_u^+} V^\dagger_{h_u^+ j} ~~~~~ -G^{R*}_{ij} = U_{iW^-} c_{W^-} U^\dagger_{W^- j} + U_{i h_d^-} c_{h_d^-} U^\dagger_{h_d^- j}
\\
\nonumber C^L_{ij} &= -V_{iW^+} N^*_{jW_3} + \frac{1}{\sqrt{2}}V_{ih_u^+} N^*_{jh_u^0} ~~~~~~~~~~~C^R_{ij} = -U^*_{iW^-} N_{jW_3} - \frac{1}{\sqrt{2}}U^*_{ih_d^-} N_{jh^0_d}
\\
D^R_{ij} &= \sin\beta V_{ih_u^+}U_{jW} + \cos\beta V_{iW^+} U_{jh_d^-}~~~~~~D^L = (D^R)^\dagger
\end{align}

A priori, in the framework we are working in, these diagrams would seem not to be important, as the gaugino masses contain no phases at the high scale so the imaginary part of the chargino and neutralino diagonalisation matrices would be zero. However, phases may be introduced by the running of the gaugino masses, given here \cite{Martin:1993zk}:

\begin{align}
\nonumber \frac{d M_a}{dt} &= \frac{2g_a^2}{16\pi^2}B_a^{(1)} M_a \\&+ \frac{2g_a^2}{(16\pi^2)^2}\left[\sum_{b=1}^3 B_{ab}^{(2)} g_b^2 (M_a + M_b) + \sum_{x=u,d,e} C_a^x \left( \Tr[Y_x^\dagger \hat{A}_x] - M_a \Tr[Y_x^\dagger Y_x]\right) \right]
\end{align}
with $B_a^{(1)}, B_{ab}^{(2)}, C_a^x$ being matrices of group coefficients, which are also found in \cite{Martin:1993zk}.

To 1st loop order there will still be no phases resulting from the running of the gaugino masses as there are no terms that would contain phases. However, at two loop order, phases can be introduced by the trilinear couplings. The term $\Tr[Y_x^\dagger Y_x]$ is manifestly real, and therefore will not contain phases. However, the term $\Tr[Y_x^\dagger \hat{A}_x]$ could well cause a phase to enter the gaugino masses at the low scale in the event where the Yukawa matrices are not aligned with the trilinears. This term would disappear in the case of alignment, as the two matrices would be diagonalised by the same left and right unitary matrices. Then we would have
\beq
\Tr[Y_x^\dagger \hat{A}_x] = \Tr[V_R Y_x^{diag} V_L^\dagger V_L \hat{A}^{diag}_x V_R^\dagger]
\eeq
which is also manifestly real.

Since generically in the M-theory framework we expect the trilinears to not be aligned with the Yukawas, we compute how large a phase one could get at the low scale given $\Order(1)$ phases in the trilinear via the Yukawa matrix, and therefore how this would enter into the expressions for the EDMs.

For this purpose we parameterise the trilinear matrix $\hat{A}_x=A_x Y_x$ where
\beq
A_x =
\bmat
A_0 e^{i\phi_1} & 0 & 0 \\
0 & A_0 e^{i\phi_2} & 0 \\
0 & 0 & A_0 e^{i\phi_3}
\emat
\eeq

To a good approximation, the dominant contribution to the phase in $\Tr[Y_x^\dagger \hat{A}_x]$ will come from the third generation phase, as the others are suppressed by powers of the small parameter $\epsilon$ in all the textures we consider here. Consequently the result is largely texture independent in these two-loop diagrams. This is due to the $1$ in the $\{3,3\}$ position of the Yukawa matrix being texture-independent, resulting in the dominance of the third generation phase. This is different from the 1-loop results, as in those diagrams, the contribution from the third generation is suppressed relative to the first generation contribution.

Thus we can do a calculation of approximately how big an imaginary part $M_a$ will have at the low scale by considering the running of the imaginary part only. For the purpose of the calculation, we work in a situation where we rotate to a basis where the down-type Yukawa matrix is diagonal, so the phases are contained in the up-type Yukawa matrix. 

\beq
\begin{split}
\frac{d \im(M_a)}{dt} &\simeq \frac{2g_a^2C_a^u}{(16\pi^2)^2}\im(\Tr[Y_u^\dagger \hat{A}_u]) \\
&\simeq \frac{2g_a^2C_a^x}{(16\pi^2)^2} A_0 \sin\phi_3 \\
\end{split}
\eeq
then we find that 
\beq
\im(M_a) \simeq \frac{2g_a^2C_a^x}{(16\pi^2)^2} A_0 \sin\phi_3 \log\left[ \frac{M_Q}{M_{GUT}}\right]
\eeq
such that
\beq
\im(M_2) \simeq -200 \sin\phi_3 \left( \frac{A_0}{75 \TeV}\right)
\eeq
where we have used \beq C_a^{u,d,e} = \bmat 26/5 & 14/5 & 18/5 \\ 6 & 6 & 4 \\ 4 & 4 & 0\emat \eeq as found in \cite{Martin:1993zk}.  This is closely comparable to the result of doing the full two loop running of the gaugino masses and the gauge couplings using the package RGERun2.0 available for Mathematica.

Knowing how large a phase can appear in the gaugino masses, we turn to computing the diagonalisation matrices for the chargino and neutralino mass matrices, as the phases on said diagonalisation matrices will appear in the expressions for the EDMs.

The chargino mass matrix is known to be
\beq
X=
\bmat
M_2 & \sqrt{2}\sin\beta M_W \\
\sqrt{2} \cos\beta M_W & \mu
\emat
\eeq
where $M_2$ is complex and the other entries are real. This matrix is diagonalised by the following rotation
\beq
X^{diag}=U^* X V^{-1}
\eeq
where U and V are unitary matrices for which analytic expressions can be obtained. We do so by solving for V and U given that
\beq
V X^\dagger X V^{-1} = U^*X X^\dagger U^T = \bmat M_{C_1}^2 & 0 \\ 0 & M_{C_2}^2 \emat
\eeq

We parameterise V and U as
\beq
U,V=
\bmat
c_{U,V} & t_{U,V} c_{U,V} \\
-t^*_{U,V} c_{U,V} & c_{U,V}
\emat
\eeq
where $c_i = \cos\theta_i$ and $t_i = \tan\theta_i$, which we solve for. 

We give here expressions for $X^\dagger X$ and $X X^\dagger$ in order to simplify our expressions for $c_i$ and $t_i$.
\beq
\begin{split}
X^\dagger X &\equiv \bmat A_{11} & A_{12} \\ A_{21} & A_{22} \emat = \bmat 
|M_2|^2 + 2 \cos\beta^2M_W^2 & M_2^*\sqrt{2}\sin\beta M_W + \sqrt{2} \cos\beta M_W \mu \\
M_2 \sqrt{2}\sin\beta M_W + \sqrt{2} \cos\beta M_W \mu & 2\sin\beta^2 M_W^2 + \mu^2
\emat
\\
X X^\dagger &\equiv \bmat B_{11} & B_{12} \\ B_{21} & B_{22} \emat = \bmat 
|M_2|^2 + 2 \sin\beta^2M_W^2 & M_2\sqrt{2}\cos\beta M_W + \sqrt{2} \sin\beta M_W \mu \\
M_2^* \sqrt{2}\cos\beta M_W + \sqrt{2} \sin\beta M_W \mu & 2\cos\beta^2 M_W^2 + \mu^2
\emat
\end{split}
\eeq

Then we find that 
\beq
\begin{split}
t_V &= \frac{(A_{22}-A_{11}) \pm \sqrt{(A_{11}-A_{22})^2 + 4A_{21}A_{12}}}{2A_{21}},~~t_V^* = {t_V}_{A_{ij}\leftrightarrow A_{ji}}
\\
t_U &= \frac{(B_{22}-B_{11}) \pm \sqrt{(B_{11}-B_{22})^2 + 4B_{21}B_{12}}}{2B_{12}},~~t_U^* = {t_U}_{B_{ij}\leftrightarrow B_{ji}}
\end{split}
\eeq
with $c_i = \frac{1}{\sqrt{1+ |t_i|^2}}$. It is noted that $A(B)_{12}$ and $A(B)_{21}$ contain the phase from the trilinears, $\phi_3$. 

This assignment for the entries of $V$ and $U$ renders X diagonal, with phases in the diagonal components. Since we want our mass eigenvalues to be real, we rotate away the phases, such that our rotation matrices are actually
\beq
\begin{split}
U' &= \bmat e^{i\phi_{C1} /2} & 0 \\ 0 & e^{i\phi_{C2}/2} \emat 
\cdot \bmat
c_{U} & t_{U} c_{U} \\
-t^*_{U} c_{U} & c_{U}
\emat
\\
V' &= \bmat e^{i\phi_{C1} /2} & 0 \\ 0 & e^{i\phi_{C2}/2} \emat 
\cdot \bmat
c_{V} & t_{V} c_{V} \\
-t^*_{V} c_{V} & c_{V}
\emat
\end{split}
\eeq
so that $U'^* X V'^{-1} = X^{RD}$ where $X^{RD}$ is real and diagonal. 

We can use these matrices to find the imaginary part which enters into the expressions for $d_f^{\gamma H}$ and $d_f^{Z H}$ through the matrices $D^{(L,R)}$ and $G^{(L,R)}$ defined as above in equation (\ref{DiagMatrices.EQ}). In order to find the imaginary part of $C^{(L,R)}$ however, we must perform a numerical diagonalisation of the neutralino mass matrix. 

We compute the electron EDM first, with 
\beq
d_e = d_e^{\gamma H} + d_e^{ZH} + d_e^{WW}
\eeq
and find that this gives us an upper bound  from equation (\ref{dGR.EQ}) of 
\beq
|d_e| <5 \times 10^{-30} e \cm
\eeq
for  $M_{\chi^+_1}\sim M_{\chi^0_2} \sim 274 \GeV, ~~M_{\chi^+_2}\sim M_{\chi^0_4} \sim 5000 \GeV$ which is well above the estimate from the leading order contributions. The heavier neutralino's mass is dominated by the $\mu$ term from the superpotential, which in the M-theory is found to be of order $0.1 m_{3/2} \sim 5000 \GeV$ \cite{Acharya:2011te}. Since this contribution is from a two loop effect, it does not depend on the scalars, but rather on the much lighter neutralinos and charginos. 

The neutron EDM upper bound from these diagrams comes from using equations (\ref{dGR.EQ}, \ref{nedm}) and is
\beq
|d_n| < 5 \times 10^{-29} e \cm
\eeq
for the same values of the chargino and neutralino masses as for the electron. Again, this is well above the estimate from the leading order contributions. The ratio of $d_n/d_e \sim 10$ is approximately in line with the results in \cite{Giudice:2005rz}. Our values are about two orders of magnitude lower than their reported results, primarily because we actually compute the phases in the diagram from the high scale, rather than taking it to be some $\Order(1)$ factor. 

These are upper limits given the predicted values of $m_{3/2}$, $M_2$ and $\mu$, and therefore could change given different input values. The misalignment of the trilinears and the Yukawas is as yet unknown, so we use $\sin\phi_3 \sim \Order(1)$ here. A precise value of $\sin\phi_3$ could in principle be determined for a given Yukawa texture given the known CKM phase and known misalignment.

These are the dominant contributions in the generic case where the trilinears are not aligned with the Yukawas. If this is the case, then they may be accessible in the next round of experiments to measure EDMs. A measurement would of course imply that the relative hierarchy of $\mu \sim 10 \times M_2$ is correct, as for different values of these two parameters, we would get a different result. This can be seen in \cite{Giudice:2005rz}. Of note is that these results are independent of the choice of texture, due to the third generation dominating. Therefore, a measurement of an EDM would not allow us to learn the high scale structure of the Yukawa textures. If the experiments were to not detect an EDM, this would suggest that either the trilinears are aligned with the Yukawa matrices, or the phases in the trilinears are indeed small. Thus, a non-detection would give us a better understanding of the relation between the full trilinear matrices and the Yukawa matrices, regardless of what texture we are considering.  

\subsection{Sub-dominant one-loop contributions}
\begin{figure}
\centering
\includegraphics[scale=0.4]{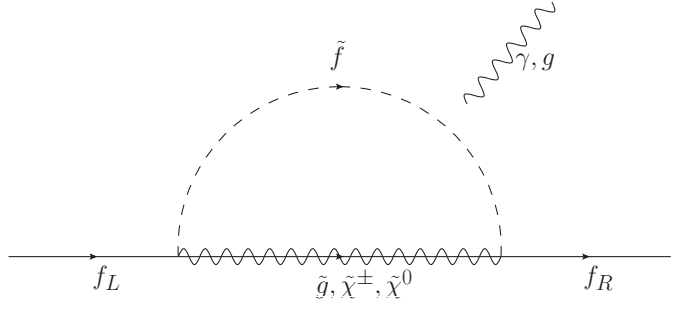}
\caption{One-loop contributions to the fermion EDMs, with scalars running in the loop.}
\label{OneLoop.FIG}
\end{figure}

In the situation where the trilinears are aligned with the Yukawas, the two loop result would be zero, as no phase would enter the gaugino masses, so the diagrams we consider above would not give a contribution to the EDMs. However, the one loop contribution would not be zero. Therefore, we consider here the five-dimensional electric and chromo-electric couplings at one-loop, as seen in Fig. \ref{OneLoop.FIG}. 

As seen in Appendix \ref{1Loop.APP}, we can express the chromo-EDM for the quarks in terms of the small ratio $r \equiv m_i^2/m_{\sq}^2$, with $i = \tilde{\chi}^0, \tilde{\chi}^\pm, \tilde{g}$. The gluino loop dominates, as also seen in Appendix \ref{1Loop.APP}, so the largest contribution comes from 
\beq
d_{q}^C \sim \frac{g_s \alpha_s}{4\pi}    \frac{\im(A^q_{SCKM})}{m_{\sg}^3} r^2 \left[ C(r) + r C^{'}(r) \right]
\eeq
The quark EDM contributions are negligible, as they depend on chargino and neutralino loops only, and thus the largest contribution will still be smaller than the quark CEDM. The term $\im(A^q_{SCKM})$ contains the phases that entered the diagonal entries from the running and subsequent diagonalsation to the super-CKM basis. The relevant results for this are given in the following table for each of the textures considered.
\begin{table}[h]
\resizebox{0.99\textwidth}{!}{\begin{minipage}{\textwidth}
\center
\begin{tabular}{c || c  c  c  c  c  c }
\hline
\textbf{Texture} & $\bm{\im({A^u_{11}})}$ & $\bm{\im({A^u_{22}})}$ & $\bm{\im({A^u_{33}})}$ & $\bm{\im({A^d_{11}})}$ & $\bm{\im({A^d_{22}})}$ & $\bm{\im({A^d_{33}})}$ \\
\hline \hline 
& & & & & & \\
\textbf{1} & $27A_0Y^u_{11}\epsilon^6$ & $18A_0 Y^u_{22}\epsilon^4$ & $9A_0Y^u_{33}\epsilon^4 $ & $3A_0Y^d_{11} \epsilon^{15}$ & $3A_0Y^d_{22} \epsilon^{10}$ & $3A_0 Y^d_{33} \epsilon^4$ \\
& & & & & & \\
\hline
& & & & & & \\
\textbf{2} & $3A_0Y^u_{11}\epsilon^{12}$ & $3A_0 Y^u_{22}\epsilon^{10}$ & $3A_0 Y^u_{33}\epsilon^4$ & $27A_0Y^u_{11}\epsilon^6$ & $18A_0 Y^u_{22}\epsilon^4$ & $9A_0 Y^u_{33}\epsilon^4$ \\
& & & & & & \\
\hline
& & & & & & \\
\textbf{3} & $12\sqrt{2} A_0  Y^u_{11}\epsilon^7$ & $60A_0 Y^u_{22}\epsilon^6$ & $48A_0  Y^u_{33}\epsilon^6$ & $72A_0 Y^d_{11}\epsilon^7$ & $144A_0 Y^d_{22}\epsilon^3$ & $288A_0 Y^d_{33}\epsilon^6$ \\
& & & & & & \\
\hline
& & & & & & \\
\textbf{4} & $12 A_0  Y^u_{11}\epsilon^7$ & $9A_0 Y^u_{22}\epsilon^2$ & $18A_0  Y^u_{33}\epsilon^4$ & $72A_0 Y^d_{11}\epsilon^7$ & $6A_0 Y^d_{22}\epsilon^2$ & $3A_0 Y^d_{33}\epsilon^4$ \\
& & & & & & \\
\hline
& & & & & & \\
\textbf{5} & $18 A_0 Y^u_{11} \epsilon^4$ & $60A_0 Y^u_{22}\epsilon^6$ & $48A_0  Y^u_{33}\epsilon^6$ & $36A_0 Y^d_{11}\epsilon^7$ & $36A_0 Y^d_{22}\epsilon^3$ & $288A_0 Y^d_{33}\epsilon^6$ \\
& & & & & & \\
\hline
& & & & & & \\
\textbf{6} & $24/\sqrt{2} A_0 Y^u_{11} \epsilon^7$ & $9A_0Y^u_{22}\epsilon $ & $18A_0  Y^u_{33}\epsilon^4$ & $36A_0 Y^d_{11}\epsilon^7$ & $3A_0 Y^d_{22}\epsilon^4$ & $3A_0 Y^d_{33}\epsilon^4$ \\
& & & & & & \\
\hline
& & & & & & \\
\textbf{7} & $9 A_0 Y^u_{11}\epsilon^4 $ & $9/\sqrt{2}A_0 Y^u_{22}\epsilon^2$ & $9A_0  Y^u_{33}\epsilon^4$ & $3/\sqrt{2}A_0 Y^d_{11}\epsilon^6$ & $3/2 A_0 Y^d_{22}\epsilon^4$ & $3/2 A_0 Y^d_{33}\epsilon^4$ \\
& & & & & & \\
\hline
\end{tabular}
\center
\end{minipage} }
\caption{The results for the various textures in terms of the diagonal Yukawa matrix elements $Y^\alpha_{ii}$. The first three columns are for up-type, and the second three are for down-type.}
\end{table}

Thus, if we make the definition 
\beq
K^\alpha_{ii} \equiv \frac{\im({A^\alpha_{ii}})}{A_0 Y^\alpha_{ii}} \sim \frac{\im({A^\alpha_{ii}})}{m_{\sq}}
\eeq
we can present the results numerically for the various values of $\epsilon$ considered. As a reminder, for textures 1 and 2, $\epsilon^u \sim 0.26,~\epsilon^d \sim 0.27$, and for textures 3-7, $\epsilon^u = \epsilon^d \sim 0.22$.

We rewrite here the expression for $d^C_q$ in such a way as to present our results more clearly for given textures. 
\beq
d_q^C \sim \frac{g_s \alpha_s}{4\pi}    \frac{K^q m_{\sg}^3}{m_{\sq}^3} \left[ C(r) + r C^{'}(r) \right] 
\eeq

\begin{table}[h]
\resizebox{0.99\textwidth}{!}{\begin{minipage}{\textwidth}
\centering
\begin{tabular}{c || c  c  c | c  c  c }
\hline
\textbf{Texture} & $\bm{K^u_{11}}$ & $\bm{K^u_{22}}$ & $\bm{K^u_{33}}$ & $\bm{K^d_{11}}$ & $\bm{K^d_{22}}$ & $\bm{K^d_{33}}$ \\
\hline \hline 
& & & & & & \\
\textbf{1} & $8 \times 10^{-3}$ & $0.08$ & $0.04$ & $9\times10^{-9}$ & $6\times 10^{-6}$ & $0.02$ \\
& & & & & & \\
\hline
& & & & & & \\
\textbf{2} & $3\times10^{-7}$ & $4\times10^{-6}$ & $0.02$ & $0.01$ & $0.1$ & $0.05$ \\
& & & & & & \\
\hline
& & & & & & \\
\textbf{3} & $4 \times 10^{-4}$ & $7\times10^{-3}$ & $5\times10^{-3}$ & $2\times10^{-3}$ & $2$ & $0.03$ \\
& & & & & & \\
\hline
& & & & & & \\
\textbf{4} & $3\times 10^{-4}$ & $0.4$ & $0.04$ & $2\times10^{-3}$ & $0.3$ & $7\times10^{-3}$ \\
& & & & & & \\
\hline
& & & & & & \\
\textbf{5} & $0.04$ & $7\times10^{-3}$ & $5\times10^{-3}$ & $9\times10^{-4}$ & $0.4$ & $0.03$ \\
& & & & & & \\
\hline
& & & & & & \\
\textbf{6} & $2\times10^{-3}$ & $2$ & $0.04$ & $9\times10^{-4}$ & $7\times10^{-3}$ & $7\times10^{-3}$ \\
& & & & & & \\
\hline
& & & & & & \\
\textbf{7} & $0.02$ & $0.3$ & $0.02$ & $2\times10^{-4}$ & $4\times10^{-3}$ & $4\times10^{-3}$ \\
& & & & & & \\
\hline
\end{tabular}
\end{minipage} }
\caption{Numerical results for the various pre-factors $K^\alpha$. The first three columns are for up-type, and the second three are for down-type.}
\end{table}

From observing this table, we see that the largest $K^\alpha$ are for the 2nd and 3rd generations. However, the appearance of these in the loop are suppressed, so only the 1st generation need be considered for calculating the upper bound on the quark contribution to the EDM of the neutron. 

Thus we present in table \ref{OneLoopResults.TAB} the upper bounds on the neutron EDM for the various textures, given our results above, and using the relation in equation (\ref{nedm}).

\begin{table}[h]
\resizebox{0.86\textwidth}{!}{\begin{minipage}{\textwidth}

\begin{tabular}{c || p{1.9cm} p{1.9cm} p{1.9cm} p{1.9cm} p{1.9cm} p{1.9cm} p{1.9cm} }
\hline
& & & & & & &\\
\scalebox{1.1}{\textbf{Texture}} & \textbf{1}  & \textbf{2}  & \textbf{3} & \textbf{4} & \textbf{5} & \textbf{6} & \textbf{7} \\
& & & & & & &\\
\hline \hline
& & & & & & &\\
\scalebox{1.2}{$\bm{d_u}$}$(\times 10^{-31})$ & $4 $ & $2 \times 10^{-5}$ & $0.2$ & $0.2 $ & $20 $ & $1$ & $10 $\\
& & & & & & &\\
\hline
& & & & & & &\\
\scalebox{1.2}{$\bm{d_d}$} $(\times 10^{-31})$& $6 \times 10^{-7}$ & $6$ & $1$ & $1$ & $0.5$ & $0.5$ & $0.1$\\
& & & & & & &\\
\hline
& & & & & & &\\
\scalebox{1.2}{$\bm{d_s}$}$(\times 10^{-31})$ & $4 \times 10^{-3}$ & $60$ & $900$ & $200$ & $200$ & $4$ & $2$\\
& & & & & & &\\
\hline
& & & & & & &\\
\scalebox{1.2}{$\bm{|d_n|}$}$(\times 10^{-31})$&$1$&$8$&$1$&$1$&$7$&$0.3$ & $4$ \\
& & & & & & &\\
\hline
& & & & & & &\\
\scalebox{1.2}{$\bm{|d_{Hg}|}$}$(\times 10^{-31})$ &$3\times 10^{-2}$&$3\times 10^{-2}$&$7\times 10^{-2}$&$9\times 10^{-3}$&$0.2$&$4\times 10^{-3}$ & $8\times 10^{-2}$ \\
& & & & & & &\\
\hline
\end{tabular}
\end{minipage} }
\caption{Results for the up, down and strange quark EDMs, and the neutron and mercury EDMs for the various textures.}
\label{OneLoopResults.TAB}
\end{table}

We see that the maximal prediction is from texture 2, which gives an upper bound for the neutron EDM of
\beq
|d_n| \sim 8 \times 10^{-31} \cdot \left(\frac{m_{\sg}}{1 \TeV}\right) \left( \frac{50 \TeV}{m_{\sq}}\right)^3 \left(\frac{\epsilon}{0.26}\right)^6 e \cm
\eeq
We remark here that this is of order $\sim 100 \times$ the expected SM result \cite{Pospelov:2013sca}. 

The maximal prediction for the mercury EDM can also be seen in table \ref{OneLoopResults.TAB}, and is given by texture 5, with an upper bound of

\beq
|d_{Hg}| \sim 2 \times 10^{-32} \cdot \left(\frac{m_{\sg}}{1 \TeV}\right) \left( \frac{50 \TeV}{m_{\sq}}\right)^3  e \cm
\eeq
We do not give an $\epsilon$ dependence for the mercury EDM as it depends on a combination of $d_u,~d_d$ and $d_s$, all of which have different $\epsilon$ dependences.

We then turn to the results for the electron EDM. From Appendix \ref{1Loop.APP}, we know that the only diagram that contributes is the neutralino exchange, since if the two-loop contribution is absent, there are no CP violating phases coming from the chargino sector in the theory due to the alignment of the Trilinears with the Yukawa matrices. Thus we have that 

\beq
d_e^E \sim \frac{e \alpha_{EM}}{4\pi \cos\theta_W} \frac{\im(\hat{A}^e_{SCKM})m_e}{m^3_{\tilde{B}}} r^2 \left[ B(r) + r B'(r)\right]
\eeq
where the variable r is defined as $r \equiv \frac{m_{\tilde{B}}^2}{m_{\tilde{e}}^2}$. We recall here that $\im(\hat{A}^e_{ii}) \sim K^e_{ii}m_{\tilde{e_i}}$, and so this can be rewritten as
\beq
d_e^E \sim \frac{e \alpha_{EM}}{4\pi \cos\theta_W}m_e m_{\tilde{B}} \frac{K^e }{m^3_{\tilde{e}}}  \left[ B(r) + r B'(r)\right]
\eeq
Our results are summarised in table \ref{ElectronOneLoop.TAB}.
\begin{table}[h]

\resizebox{0.88\textwidth}{!}{\begin{minipage}{\textwidth}

\begin{tabular}{p{2.5cm} || p{1.6cm}  p{1.9cm}  p{1.9cm}  p{1.9cm}  p{1.9cm}  p{1.9cm}  p{1.9cm} }
\hline
& & & & & & &\\
\scalebox{1.1}{\textbf{Texture}} & \textbf{1}  & \textbf{2} & \textbf{3} & \textbf{4} & \textbf{5} & \textbf{6} & \textbf{7} \\
& & & & & & &\\
\hline \hline
& & & & & & &\\
\scalebox{1.2}{$\bm{d_e}$} $(\times 10^{-34})$& $5 $  & $2\times 10^{-3}$ & $0.2$ & $0.2$ & $0.2$ & $0.2$ & $5 \times 10^{-2}$\\
& & & & & & &\\
\hline
\end{tabular}
\end{minipage} }
\caption{Results for the electron EDM for the various textures.}
\label{ElectronOneLoop.TAB}
\end{table}

We see that the maximal prediction is 
\beq
d_e \sim 5 \times 10^{-34} \left(\frac{m_{\tilde{B}}}{200 \GeV}\right) \left( \frac{50 \TeV}{m_{\sq}}\right)^3 \left( \frac{\epsilon}{0.36}\right)^6e \cm
\eeq
as a result of using texture 1. We remark here that this is of order $10^5 \times$ the SM prediction \cite{Pospelov:2013sca}. 

While texture 2 gave the highest neutron EDM, it actually gives the smallest electron EDM. This comes about primarily as a result of two factors. The first is to do with the naive dimensional analysis approach to calculating the EDM, in that the EDM of the down quark contributes $4/3$ whereas the up quark contributes $-1/3$. Thus despite the larger value of $K^u_{11}$ in texture 5, the slightly smaller value of $K^d_{11}$ in texture 2 gives a larger result, albeit marginally. This also explains why texture 7, despite giving the largest $d_u$, ends up giving a slightly smaller $d_n$, as the $d_u$ contribution loses out to the $d_d$ part.

The second reason, which applies to both texture 5 and texture 2, is due to the running. The largest contributions to the EDM arise when the smallest powers of $\epsilon$ from running coincide with the largest powers of $\epsilon$ in the eigenvalues. In this case, in texture 2, $Y^d_{11} \propto \epsilon^5$, and the running contributed a factor of $27\epsilon^6$. In texture 5, $Y^u_{11} \propto \epsilon^4$, and the running contributed a factor of $18\epsilon^4$. The pre factors from the running of the full trilinears are important because they help counteract the large powers of $\epsilon$.

The electron EDM is suppressed relative to the neutron EDM for a few reasons. Chief among them is that we have factors of $\alpha_{EM}$ rather than $\alpha_s$, due to the electroweak nature of the diagram. Another suppression arises due to the loop factor in the electron EDM diagram, $B(r) + rB'(r)$ being substantially smaller than the loop factor in the gluino exchange diagram for the quark CEDMs.

It is curious that texture 2, while giving the largest neutron EDM results in the smallest electron EDM. This comes about because in textures 1 and 2, we assume that the Yukawa texture for the leptons is of the same form as that of the up quarks. Therefore when we start in a basis where the up Yukawa texture is diagonal, while it does pick up factors from the running, they are typically large powers of $\epsilon$.

%%%%%%%%%%%%%%%%%%CONCLUSION%%%%%%%%%%%%%%%%%%%%%%

\section{Conclusion}
\label{Conclusion.SEC}

\begin{table}[h]
\label{comparison.TAB}
\resizebox{0.85\textwidth}{!}{\begin{minipage}{\textwidth}
\begin{tabular}{c || c c c c c c c}
\hline
 & \textbf{\small{M-Theory}} & \textbf{\small{M-Theory}} & \textbf{\small{Split SUSY}} & \textbf{\small{Generic SUSY}} & \textbf{\small{Current Limit}}Ê& \textbf{\small{SM value}}\\
 & \scriptsize{(2-loop) } & \scriptsize{(1-loop)} & \scriptsize{($\mu \sim \frac{m_{3/2}}{10},~M_2 \sim \frac{1}{3} \TeV$)} &  & & \\
 \hline
 $\bm{d_e}$ \scriptsize{($\times 10^{-28}~e \cm$)} & $5 \times 10^{-2}$ & $5 \times 10^{-6}$ & $\sim1$ & $\sim 1000$ & 0.87 & $10^{-10}$\\
 $\bm{d_n}$ \scriptsize{($\times 10^{-28}~e \cm$)} & $0.5$ & $8 \times 10^{-3}$ & $\sim 10$ & $\sim 1000$ & 300 & $10^{-4}$\\
 $\bm{d_{Hg}}$ \scriptsize{($\times 10^{-28}~e \cm$)}& $5 \times 10^{-4}$ & $2 \times 10^{-4}$ & N/A & N/A & 0.31 & N/A\\
 $\bm{d_n/d_e}$ & $\sim 10$ & $\sim 10^3$ & $\sim 9$ & N/A & N/A & $\sim10^6$\\
 \hline
\end{tabular}
\end{minipage} }
\caption{Results for the possibly dominant two-loop and 1-loop predictions of EDMs from compactified M-Theory, as compared to the predictions from Split SUSY \cite{ ArkaniHamed:2004yi, Giudice:2005rz}, Generic SUSY models \cite{Pospelov:2005pr}, the current limits \cite{Baron:2013eja, Baker:2006ts,Griffith:2009zz} and the expected SM value \cite{Pospelov:2005pr}.}
\end{table}

In this paper we have discussed how the CP-violating phases in compactified M-theory arise only in the Yukawa sector at the high scale, but nevertheless give rise to low scale EDMs via RGE running and the Super-CKM rotation. Therefore there will be a dependence on the Yukawa textures at the high scale. For various textures the running and subsequent diagonalisation of the full trilinear couplings $\hat{A}_{\alpha \beta \gamma}$ to the Super-CKM basis causes them to pick up phases at the low scale.

We have estimated the electron, neutron and mercury EDMs for textures at the high scale, all of which satisfy experimental constraints on quark masses and CKM matrix elements. The dominant source of EDMs in the generic case where the trilinears are not aligned with the Yukawa matrices are from two-loop diagrams involving charginos and neutralinos, and are therefore not suppressed by large scalar masses. These contributions are much larger than the one-loop diagrams as a result. While at the high scale the phase in the gaugino masses is zero, a misalignment between the trilinears and the Yukawas induces a non-zero phase at two-loop order in the running of the masses. Thus there is a non-zero phase at the low scale. 

A priori one would think that a phase that arises only at two-loop order, which is then inserted into a two-loop effect, would be smaller than the one-loop contribution. However, several factors contribute to making the two-loop effects large. The phase induced by the running of the gaugino masses depends on the full trilinear, which is large ($\hat{A} \sim \Order(75 \TeV)$) in the M-theory compactification, which results in a relatively large phase, despite the two-loop suppression. Further, the two-loop contribution is approximately $d \sim (\alpha^2/\pi^2) (m_f m_{\sch} / M_{EW}^3)$, as opposed to the $d \sim (\alpha_s^2/\pi) (m_f m_{\sg} /m_{\sq}^3 )$ for the one-loop contribution. Since $m_{\sq}^3 \gg m_{EW}^3$, the two-loop contribution turns out to be quite large when the phases are not small. 

These two-loop contributions do not depend on the choice of texture, as they arise mainly due to the third generation phase entering the gaugino mass running. Since the third generation coupling is always 1 in the textures we consider here, the texture-dependence is negligible.

We summarise our results and compare with other models in table \ref{comparison.TAB}. As seen there, the estimated upper bounds we find from the two loop contributions are $|d_n| \lesssim 5 \times 10^{-29} e \cm$ and $d_e \lesssim 5 \times 10^{-30}e \cm$. These are values that are likely to be accessible in the near future. A detection would confirm a misalignment between the trilinears in the Soft Lagrangian and the Yukawa matrices. Non-detection would imply that they are aligned, or that the phase in the trilinears is indeed small. The results are different from those reported in the Split SUSY scenario \cite{ArkaniHamed:2004yi, Giudice:2005rz}, as can be seen in table \ref{comparison.TAB} and would therefore provide a means of distinguishing between that scenario and the compactified M-theory. Further, the ratio of the EDM predictions from the two-loop diagrams is a strong test of the $M_2/\mu$ ratio predicted in the compactified M-theory. These results assume the strong CP contribution is small. 

We also compute the sub-dominant one-loop results. The reason being that these would provide the dominant contributions in the case where the trilinears are indeed aligned with the Yukawas. In this case, the estimated upper bounds turn out to depend strongly on the textures and are all below current experimental limits. We argue that although we only study some textures, the results for EDM upper limits are generic. We find that the Electroweak contribution to the neutron and mercury EDMs is larger than the expected strong contribution in the SM, so any observation of a neutron or mercury EDM may be interpreted as the Electroweak part, but this would require further study. The upper bound we estimate for the electron EDM is well below current experimental limits, so we do not expect experiments in the near future to be able to measure a non-zero EDM. 

These results, while done in the context of a compactified M-theory, are likely to be applicable for supersymmetric models which have scalars similar to the M-theory ones (and would scale as the scalar mass cubed), and light gauginos, with CP-violating phases arising only in the Yukawa sector at the high scale.  

The upper bounds we find for $d_n \lesssim 8 \times 10^{-31} e \cm$, $d_{Hg} \lesssim 2 \times 10^{-32} e\cm$ and $d_e \lesssim 5 \times 10^{-34}e \cm$ are strong and testable predictions of compactified M-theory. They are much smaller than the sizes expected in supersymmetric and other generic models, but still significantly larger than the SM predictions. Unfortunately, in this case, where the trilinears are aligned with the Yukawas, we expect that non-zero EDMs will not be found until there are major improvements in experimental sensitivity.

The next round of experimental measurements of EDMs will provide valuable insight into the fundamental Yukawa couplings of the quarks and leptons. If non-zero EDMs are measured, it would suggest that there is indeed a misalignment between the full trilinears and the Yukawa couplings, with the dominant EDM contribution arising at two-loop order. If non-zero EDMs are not found, it would suggest that there is alignment between the trilinears and the Yukawas.  Further advances in experimental sensitivity might then provide some insight into the structure of the Yukawas at the high scale, given the strong texture dependence of the dominant one-loop contributions.

\section*{\small{Acknowledgements}}

We would like to thank John Ellis, Piyush Kumar and Bob Zheng for useful discussions. The work of SARE and GLK is supported in part by Department of Energy grant DE-SC0007859.
%%%%%%%%%%%%%%%%%%APPENDIX%%%%%%%%%%%%%%%%%%%%%%%%

%%%%%%%%%%%%%%%%%%YUKAWA TEXTURES%%%%%%%%%%%%%%%%%%
\begin{appendices}

\section{Yukawa texture derivations}
\subsection{Textures where one matrix is diagonal}

In the special case where we can rotate to a basis where one of the up or down quark Yukawa matrices is diagonal, the derivation of the other Yukawa matrix is greatly simplified, since $V_{CKM} = (U^L_u)^\dagger U^L_d$ depends on the diagonalisation matrices.

In the case where the down Yukawa matrix is taken to be diagonal, this simplifies to $V_{CKM} = (U^L_u)^\dagger  \mathbb{1}_{3\times3}$. Then since we know that $(U^L_u)^\dagger Y^u (Y^u)^\dagger U^L_u = M_u^2$, where $M_u^2$ is the diagonal matrix of the quark masses squared, we can solve for $Y^u$ using the expression $Y^u (Y^u)^\dagger = U^L_u M^2_u (U^L_u)^\dagger$. The expression for this is given below:

\beq
\begin{split}
Y^u (Y^u)^\dagger &= 
\bmat
m_u^2 + \epsilon^2 m_c^2 + \epsilon^6 m_t^2 & \epsilon m_u^2 + \epsilon m_c^2 + \epsilon^5 m_t^2 & \epsilon^3m_u^2 + \epsilon^3 m_c^2 + \epsilon^3 m_t^2 \\
\epsilon m_u^2 + \epsilon m_c^2 + \epsilon^5 m_t^2 & \epsilon^2 m_u^2 +  m_c^2 + \epsilon^5 m_t^2 & \epsilon^4 m_u^2 + \epsilon^2 m_c^2 + \epsilon^2 m_t^2 \\
\epsilon^3m_u^2 + \epsilon^3 m_c^2 + \epsilon^3 m_t^2 & \epsilon^4 m_u^2 + \epsilon^2 m_c^2 + \epsilon^2 m_t^2 &  \epsilon^6m_u^2 + \epsilon^4 m_c^2 + m_t^2 
\emat
\\
&\approx
\bmat
\epsilon^{16} + \epsilon^{10} + \epsilon^6 & \epsilon^{17} + \epsilon^{9} + \epsilon^5 & \epsilon^{19} + \epsilon^{11} + \epsilon^3 \\
\epsilon^{17} + \epsilon^{9} + \epsilon^5 & \epsilon^{18} + \epsilon^{8} + \epsilon^4 & \epsilon^{20} + \epsilon^{10} + \epsilon^2 \\
\epsilon^{19} + \epsilon^{11} + \epsilon^3 & \epsilon^{20} + \epsilon^{10} + \epsilon^2 & \epsilon^{22} + \epsilon^{12} + 1
\emat
\end{split}
\eeq
where we have used the approximate hierarchy as described in equation (\ref{hierarchy.EQ}). We then use the ansatz 
\beq
Y^u = \bmat \epsilon^8 & \epsilon^a & \epsilon^b \\ \epsilon^c & \epsilon^4 & \epsilon^d \\ \epsilon^e & \epsilon^f & 1\emat
\eeq
and solve for $a,~b,~c,~d,~e,~f$.

The same analysis can be repeated for the case where the up-type Yukawa matrix is diagonal. 

\subsection{Minimal matrix derivation}

It is of interest to consider what the minimal matrix would be, and whether it is symmetric or not. The only assumption we start with in this derivation is that we know the hierarchy, which is the same as previously, and that the CKM matrix is parameterised by equation (\ref{CKMparam.EQ}). We know the general structure of the unitary diagonalisation matrices to be

\begin{align}
U^{L,R}_u = \bmat 1 & \epsilon^i & \epsilon^j \\ \epsilon^i & 1 & \epsilon^k \\ \epsilon^j & \epsilon^k & 1 \emat, ~~ÊU^{L,R}_d = \bmat 1 & \epsilon^x & \epsilon^y \\ \epsilon^x & 1 & \epsilon^z \\ \epsilon^y & \epsilon^z & 1 \emat
\end{align}
such that
\begin{align}
V_{CKM} = \bmat 1 & (\epsilon^i + \epsilon^x + \epsilon^{j+z}) & (\epsilon^j + \epsilon^y + \epsilon^{i+z}) \\ (\epsilon^i + \epsilon^x + \epsilon^{j+z}) & 1 & (\epsilon^k + \epsilon^z + \epsilon^{i+y}) \\ (\epsilon^j + \epsilon^y + \epsilon^{i+z}) & (\epsilon^k + \epsilon^z + \epsilon^{i+y}) & 1\emat
\end{align}
which leads to the following constraints:
\begin{itemize}
\item $i$, $x$ or $j+z$ $=1$
\item $k$, $z$ or $i+y$ $=2$
\item $j$, $y$ or $i+z$ $=3$
\end{itemize}

In the subsequent analysis, we assume that the down diagonalisation matrix has the form of the CKM matrix, i.e. $x=1,~y=3,~z=2$ in order to have our up type diagonalisation matrix unconstrained, so that to leading order the CKM matrix is satisfied.

We define the following form for the general up type Yukawa matrix

\begin{align}
Y^u = \bmat \epsilon^8 & \epsilon^l & \epsilon^m \\ \epsilon^n & \epsilon^4 & \epsilon^o \\ \epsilon^p & \epsilon^q & 1 \emat
\end{align}
such that
\begin{align}
Y^u {Y^u}^\dagger = 
\bmat 
(\epsilon^{16} + \epsilon^{2l} + \epsilon^{2m}) & (\epsilon^{4+l} + \epsilon^{8+n} + \epsilon^{m+o} ) & (\epsilon^m + \epsilon^{8+p} + \epsilon^{l+q}) \\
(\epsilon^{4+l} + \epsilon^{8+n} + \epsilon^{m+o} ) & (\epsilon^{8} + \epsilon^{2n} + \epsilon^{2o}) & (\epsilon^o + \epsilon^{4+q} + \epsilon^{n+p}) \\
(\epsilon^m + \epsilon^{8+p} + \epsilon^{l+q})  & (\epsilon^o + \epsilon^{4+q} + \epsilon^{n+p}) & (1 + \epsilon^{2p} + \epsilon^{2q})
\emat
\end{align}

We derive conditions on the variables by looking at the diagonal components of the diagonalised matrices i) ${U^L_u}^\dagger Y^u {Y^u}^\dagger U_u^L$, and ii) ${U^L_u}^\dagger Y^u U_u^R$. These allow us to determine the minimal up-type Yukawa texture to be:

\begin{align}
Y^u = 
\bmat
\epsilon^8 & \epsilon^4 & \epsilon^4 \\
\epsilon^4 & \epsilon^4 & \epsilon^2 \\
\epsilon^4 & \epsilon^q & 1
\emat
\end{align}
with $q$ unconstrained by the diagonalisation and the unitary matrix that diagonalises $Y^u$ is given by
\begin{align}
U^{L,R}_u =
\bmat
1 & \epsilon^4 & \epsilon^4 \\
\epsilon^4 & 1 & \epsilon^4 \\
\epsilon^4 & \epsilon^4 & 1
\emat
\end{align}

%%%%%%%%%%%%%%%%%%%%%%EDMS%%%%%%%%%%%%%%%%%%%

\section{Contributions to EDMs}
\subsection{One-loop SUSY contributions to EDMs}
\label{1Loop.APP}
In this subsection we present the one-loop SUSY contributions due to the Feynman diagrams in Fig. \ref{OneLoop.FIG}. We use the results of Ibrahim and Nath \cite{Ibrahim:1997gj}.

The electromagnetic contributions to fermion EDMs are as follows

\beq
\begin{split}
&\left(\frac{d^{\tilde{g}}_q}{e}\right)^{(E)} = \frac{-2\alpha_s}{3\pi} \sum^2_{k=1} \im(\Gamma^{1k}_q) \frac{m_{\tilde{g}}}{M^2_{\tilde{q}_k}} Q_{\tilde{q}} ~ B\left(\frac{m^2_{\tilde{g}}}{M^2_{\tilde{q}_k}} \right)
\\
&\left(\frac{d^{\chi^\pm}_u}{e}\right)^{(E)} = \frac{-\alpha_{EM}}{4\pi \sin^2\theta_W} \sum^2_{k=1}  \sum^2_{i=1} \im(\Gamma_{uik}) \frac{m_{\chi^\pm}}{M^2_{\tilde{d}_k}} \left[ Q_{\tilde{d}} ~B\left(\frac{m^2_{\chi^\pm}}{M^2_{\tilde{d}_k}} \right) + (Q_u - Q_{\tilde{d}}) ~A\left( \frac{m^2_{\chi^\pm}}{M^2_{\tilde{d}_k}} \right) \right]
\\
&\left(\frac{d^{\chi^\pm}_d}{e}\right)^{(E)} = \frac{-\alpha_{EM}}{4\pi \sin^2\theta_W} \sum^2_{k=1}  \sum^2_{i=1} \im(\Gamma_{dik}) \frac{m_{\chi^\pm}}{M^2_{\tilde{u}_k}} \left[ Q_{\tilde{u}} ~B\left(\frac{m^2_{\chi^\pm}}{M^2_{\tilde{u}_k}} \right) + (Q_d - Q_{\tilde{u}}) ~A\left( \frac{m^2_{\chi^\pm}}{M^2_{\tilde{u}_k}} \right) \right]
\\
&\left(\frac{d^{\chi^\pm}_e}{e}\right)^{(E)} =\frac{\alpha_{EM}}{4\pi \sin^2\theta_W} \sum^2_{i=1}  \im(\Gamma_{ei})\frac{m_{\chi^\pm}}{m^2_{\tilde{\nu}}}~A\left( \frac{m^2_{\chi^\pm}}{m^2_{\tilde{\nu}}} \right)
\\
&\left(\frac{d^{\chi^0}_f}{e}\right)^{(E)} =\frac{\alpha_{EM}}{4\pi \sin^2\theta_W} \sum^2_{k=1}\sum^4_{i=1}  \im(\eta_{fik}) \frac{m_{\chi^0}}{M^2_{\tilde{f}_k}}Q_{\tilde{f}}~B\left( \frac{m^2_{\chi^0}}{M^2_{\tilde{f}_k}} \right)
\end{split}
\eeq
where 
\beq
\Gamma^{1k}_q = e^{-i\phi_3}D_{q2k}D^*_{q1k}
\eeq
with $\phi_3$ being the gluino phase, which in our theory can be rotated away, and $D_q$ defined as $D^\dagger_q M^2_{\tilde{q}}D_q = \diag(M^2_{\tilde{q}_1}, M^2_{\tilde{q}_2})$. With the sfermion mass matrix $M^2_{\tilde{f}}$ given by

\beq
M_{\tilde{f}}^2 =
\bmat
M_L^2 + m_f^2 + M_z^2(\frac{1}{2} - Q_f \sin^2\theta_W) \cos 2\beta & m_f(A^*_f - \mu R_f) \\
m_f(A_f -\mu^* R_f) & M_R^2 + m_f^2 + M_z^2Q_f \sin^2 \theta_W \cos 2\beta
\emat
\eeq
where $R_f = \cot \beta~ (\tan \beta)$ for $I_3 = 1/2 ~(-1/2)$. The chargino vertices are given by 
\beq
\begin{split}
&\Gamma_{uik} = \kappa_u V^*_{i2}D_{d1k}(U^*_{i1}D^*_{d1k} - \kappa_d U^*_{i2}D^*_{d2k})
\\
&\Gamma_{dik} = \kappa_d U^*_{i2}D_{u1k}(V^*_{i1}D^*_{u1k} - \kappa_u V^*_{i2}D^*_{u2k})
\end{split}
\eeq
and 
\beq
\Gamma_{ei} = (\kappa_e U^*_{i2}V^*_{i1})
\eeq
where in each case $\kappa_f$ is the Yukawa coupling, defined as $\kappa_u = \frac{m_u}{\sqrt{2}m_W \sin\beta}$ and $\kappa_{d,e} = \frac{m_{d,e}}{\sqrt{2}m_W \cos\beta}$, and U and V are the unitary matrices diagonalizing the chargino mass matrix. The neutralino vertex is defined as

\beq
\begin{split}
\eta_{fik} &= \left[-\sqrt{2}\{\tan\theta_W(Q_f - I_{3_f})X_{1i} + I_{3_f}X_{2i}\}D^*_{f1k} -\kappa_f X_{bi}D^*_{f2k}\right]
\\
& \times \left[ \sqrt{2}\tan\theta_W Q_f X_{1i}D_{f2k} - \kappa_f X_{bi} D_{f1k} \right]
\end{split}
\eeq
with $b=3~~(4)$ for $I_3 = -1/2 ~~(1/2)$, and X being the unitary matrix diagonalizing the neutralino mass matrix. 
The CEDM contributions are given by
\beq
\begin{split}
\label{chromoEDMs.EQ}
& d_q^{\sg~(C)} = \frac{g_s\alpha_s}{4\pi} \sum^2_{k=1} \im(\Gamma^{1k}_q) \frac{m_{\sg}}{M^2_{\sq_k}} C \left( \frac{m^2_{\sg}}{M^2_{\sq_k}} \right)
\\
& d_q^{\sch~(C)} = \frac{-g^2 g_s}{16\pi^2} \sum^2_{k=1} \sum^2_{i=1} \im(\Gamma_{qik}) \frac{m_{\sch_i}}{M^2_{\sq_k}} B \left( \frac{m^2_{\sch_i}}{M^2_{\sq_k}} \right)
\\
& d_q^{\snu~(C)} = \frac{g^2 g_s}{16\pi^2} \sum^2_{k=1} \sum^4_{i=1} \im(\eta_{qik}) \frac{m_{\snu_i}}{M^2_{\sq_k}} B \left( \frac{m^2_{\snu_i}}{M^2_{\sq_k}} \right)
\end{split}
\eeq
And the dimension-6 Weinberg operator gives a contribution
\beq
d^G = -3\alpha_sm_t \left(\frac{g_s}{4\pi}\right)^3 \im(\Gamma^{12}_t) \frac{z_1 - z_2}{m^3_{\sg}} H(z_1, z_2, z_3) + (t \rightarrow b)
\eeq
with $z_i = \left(\frac{M_{\tilde{t}_i}}{m_{\sg}}\right)^2$, and $z_t = \left(\frac{m_t}{m_{\sg}}\right)^2$, with the two-loop function $H(z_1, z_2, z_t)$ being given by 

\beq
H(z_1, z_2, z_t) = \frac{1}{2} \int^1_0 dx \int^1_0 du \int^1_0 dy x(1-x) u \frac{N_1 N_2}{D^4}
\eeq
where
\beq
\begin{split}
&N_1 = u(1-x) + z_t x(1-x)(1-u) - 2ux[z_1y + z_2(1-y)]
\\
&N_2 = (1-x)^2(1-u)^2 + u^2 -\frac{1}{9}x^2(1-u)^2
\\
&D=u(1-x) + z_t x(1-x)(1-u) + ux[z_1y + z_2(1-y)]
\end{split}
\eeq
However, for the purpose of this analysis in this framework, the contribution from this two-loop effect is negligible, so it will not be calculated. Thus recording these equations is merely for book-keeping purposes. We consider other two-loop effects which give larger contributions in the text. Another two-loop effect is from the Barr-Zee diagram with scalars in a loop, which is treated in the next subsection. 

The functions $A(r),~ B(r)$ and $C(r)$ used in the equations above are the one-loop functions, and are given by 
\beq
\begin{split}
&A(r) = \frac{1}{2(1-r)^2} \left(3-r+\frac{2\ln r}{1-r} \right)
\\
&B(r) = \frac{1}{2(r-1)^2} \left(1+r+\frac{2 r\ln r}{1-r} \right)
\\
&C(r) = \frac{1}{6(r-1)^2} \left(10r-26+\frac{2r\ln r}{1-r} - \frac{18\ln r}{1-r}\right)
\end{split}
\eeq

The above equations are rather intractable, and in fact a number of approximations can be utilized to simplify the calculation. For example, for the neutron and mercury, the quark CEDM contributions are much larger than the quark EDM contributions (as seen in \cite{Kane:2009kv} ), so let us take the example of the dominant gluino contribution, $d_q^{\sg~(C)}$. Expanding the relevant line in equation (\ref{chromoEDMs.EQ}), and defining $r_i = \frac{m_{\sg}^2}{m_{\sq_i}^2}$, we find that

\beq
\begin{split}
\label{dqC.EQ}
d_q^{\sg~(C)} = \frac{g_s \alpha_s}{4\pi}  \left[ \im(\Gamma^{11}_q) \frac{m_{\sg}}{m_{\sq_1}^2} C(r_1) - \im(\Gamma^{12}_q) \frac{m_{\sg}}{m_{\sq_2}^2} C(r_2) \right] 
\end{split}
\eeq
But since $\Gamma^{11}_q = - \Gamma^{12}_q$, we can then simplify this further. Also $m^2_{\sq_i} = m^2_{\sq} \pm \Delta_m$, where $\Delta_m = (m^2_{\sq})_{LR}$, i.e. it is the contribution from the off-diagonal components of the squark mass matrix. We will utilize the assumption that since we are interested in the first generation squarks, the mass splitting is small compared to the squark mass.

This allows us to expand the various factors and functions above and simplify (\ref{dqC.EQ}) to the following form

\beq
\begin{split}
%d_q^{\sg~(C)} &\approx  \frac{g_s \alpha_s}{4\pi}  \im(\Gamma^{11}_q)  \frac{m_{\sg}}{m_{\sq}^2} \left\{\left(1 + \frac{\Delta_m}{m_{\sq}^2}\right) \left[ C(r) + r \frac{\Delta_m}{m_{\sq}^2} C^{'} (r)\right] -  \left(1 - \frac{\Delta_m}{m_{\sq}^2} \right) \left[ C(r) - r \frac{\Delta_m}{m_{\sq}^2} C^{'} (r) \right] \right\}
%\\
%d_q^{\sg~(C)} &\approx  \frac{2 g_s \alpha_s}{4\pi}  \im(\Gamma^{11}_q)  \frac{m_{\sg}}{m_{\sq}^2} \frac{\Delta_m}{m_{\sq}^2} \left[ C(r) + r C^{'}(r) \right]
%\\
d_q^{\sg~(C)} &\approx \frac{g_s \alpha_s}{4\pi}    \frac{\im(m_{\sq}^2)_{LR}}{m_{\sg}^3} r^2 \left[ C(r) + r C^{'}(r) \right]
\end{split}
\eeq
as was found in \cite{Kane:2009kv}. 

For the electron EDM, there are a few things we notice which simplify the calculation. The chargino component depends on $\im(\Gamma_{ei})$, which is zero in the framework considered due to the absence of CP-violating phases in the chargino sector when the trilinears and Yukawas are aligned. Thus only the neutralino diagrams contribute. If we assume no mixing, by a similar analysis to the one performed for the gluino contribution to the quark CEDM, we find that the result for the electron is given by

\beq
d_e^E \approx \frac{e\alpha_{EM}}{4\pi\cos^2\theta_W} \frac{\im(m_{\tilde{e}}^2)_{LR}}{m_{\tilde{B}}^3} r^2 \left[ B(r_1) + r_1 B^{'}(r_1)\right]
\eeq

%%%%%%%%%%%%%%%%%BARR-ZEE%%%%%%%%%%%%%%%%%%%%%%

\subsection{Barr-Zee diagram contributions}

In general, Barr-Zee type diagrams can involve squarks, charginos or neutralinos in the inner loop, with gauge bosons and or higgs bosons in the outer loop. The two-loop diagrams when the Trilinears are not aligned with the Yukawas are considered above. So here we present the two-loop results when they are aligned. In this case, since the only phases come from the Yukawa sector, we need only consider the diagrams with squarks running in the inner loop. We are particularly interested in diagrams with third generation squarks, $\tilde{t}$ and $\tilde{b}$ running in the inner loop, since they are lighter and are not suppressed by factors of $\epsilon$. The general EDM and CEDM contributions are given by

\beq
\begin{split}
\label{dfBZ.EQ}
\left(\frac{d_f^E}{e} \right) &= Q_f \frac{3\alpha_{em}}{32\pi^3} \frac{R_f m_f}{M_A^2} \sum_{q=t,b} \xi_q Q_q^2 \left[F\left( \frac{M^2_{\sq_1}}{M_A^2} \right) - F\left( \frac{M_{\sq_2}^2}{M_A^2}\right) \right]
\\
 d_f^C &= \frac{g_s \alpha_s}{64 \pi^3} \frac{R_f m_f}{M_A^2} \sum_{q=t,b} \xi_q \left[F\left( \frac{M^2_{\sq_1}}{M_A^2} \right) - F\left( \frac{M_{\sq_2}^2}{M_A^2}\right) \right]
\end{split}
\eeq
where $M_A$ is the mass of the pseudoscalar Higgs boson $A_0$, and $R_f = \cot\beta ~Ê(\tan\beta)$ for $I_3 = 1/2 ~(-1/2)$ and $F(r)$ is the two-loop function defined as

\beq
F(r) = \int^1_0 dx \frac{x(1-x)}{r-x(1-x)} \ln\left[\frac{x(1-x)}{r}\right]
\eeq
The CP-violating couplings are $\xi_{t,b}$, defined as
\beq
\begin{split}
\label{CPcouplings.EQ}
& \xi_t = -\frac{\sin2\theta_{\tilde{t}} m_t \im(\mu e^{i\delta_{t}})}{2v^2\sin^2\beta}
\\
& \xi_b= -\frac{\sin2\theta_{\tilde{b}} m_b \im(A_b e^{-i\delta_{b}})}{2v^2\sin\beta\cos\beta}
\end{split}
\eeq
where one should be careful to note that $v=246/\sqrt{2} $ GeV, and the minus signs are chosen by convention, differing from \cite{Chang:1998uc} and the associated erratum. The variables $\theta_{\tilde{t},\tilde{b}}$ are the stop and sbottom mixing angles, and $\delta_q = \arg(A_q + R_q \mu^*)$. The squark sector mixing angle is defined as

\beq
\begin{split}
\tan(2\theta_q) &= -\frac{2 m_q |\mu R_q + A_q^* |}{ M_{\tilde{Q}}^2 - M_{\sq}^2 + \cos 2\beta M_Z^2 (T_z^q - 2e_q s_w^2)}
\\
& \approx -\frac{2 m_q |\mu R_q + A_q^* |}{ M_{\tilde{Q}}^2 - M_{\sq}^2}
\end{split}
\eeq
such that we can rewrite the CP-violating couplings given in (\ref{CPcouplings.EQ}) as 

\beq
\begin{split}
\label{xi.EQ}
&\xi_t \approx \frac{y_t^2 |A^*_t + \mu \cot \betaÊ| \im(\mu e^{i\delta_t})}{M_{\tilde{Q}}^2 - M_{\tilde{t}}^2}
\\
& \xi_b \approx  \cot\beta \frac{y_b^2 |A^*_b + \mu \tan \betaÊ| \im(A_b e^{i\delta_b})}{M_{\tilde{Q}}^2 - M_{\tilde{b}}^2}
\end{split}
\eeq
which can then be used to simplify (\ref{dfBZ.EQ}) such that it reads

\begin{align}
\centering
\nonumber \left(\frac{d_f^E}{e} \right) &= Q_f \frac{3\alpha_{em}}{32\pi^3} \frac{R_f m_f}{M_A^4} \im \left[ \frac{4y_t^2}{9} \mu(A_t + \mu^* \cot \beta) F^\prime \left( \frac{M^2_{\sq_1}}{M_A^2} \right) + \frac{y_b^2}{9} A_b (A^*_b + \mu \tan \beta) \cot \beta F^\prime \left( \frac{M_{\sq_2}^2}{M_A^2}\right) \right]
\\
 d_f^C &= \frac{g_s \alpha_s}{64 \pi^3} \frac{R_f m_f}{M_A^4} \im  \left[y_t^2 \mu (A_t + \mu^* \cot \beta )F^\prime \left( \frac{M^2_{\sq_1}}{M_A^2} \right) + y_b^2 A_b (A^*_b + \mu \tan \beta) \cot \beta F^\prime \left( \frac{M_{\sq_2}^2}{M_A^2}\right) \right]
\end{align}
where in (\ref{xi.EQ}) above, $m_{\tilde{t},\tilde{b}}$ are the average masses of the stops and sbottoms respectively. In the above equations, the mass of the CP-odd Higgs mass is given by

\beq
M_A^2 = m_{H_u}^2 + m_{H_d}^2 + 2|\mu|^2
\eeq
where the first two contributions are considerably larger than that of $\mu$, but we include the $\mu$ term for completeness.

Barr-Zee contributions turn out to be very small, and are therefore not included in our final computation of $d_e$.

\end{appendices}


\newpage

\begin{thebibliography}{15}


%\cite{Pospelov:2005pr}
\bibitem{Pospelov:2005pr}
  M.~Pospelov and A.~Ritz,
  %``Electric dipole moments as probes of new physics,''
  Annals Phys.\  {\bf 318} (2005) 119
  [hep-ph/0504231].
  %%CITATION = HEP-PH/0504231;%%
  %294 citations counted in INSPIRE as of 08 Apr 2014

%\cite{Nilles:1983ge}
\bibitem{Nilles:1983ge}
  H.~P.~Nilles,
  %``Supersymmetry, Supergravity and Particle Physics,''
  Phys.\ Rept.\  {\bf 110} (1984) 1.
  %%CITATION = PRPLC,110,1;%%
  %4286 citations counted in INSPIRE as of 29 Apr 2014
  
 %\cite{Haber:1984rc}
\bibitem{Haber:1984rc}
  H.~E.~Haber and G.~L.~Kane,
  %``The Search for Supersymmetry: Probing Physics Beyond the Standard Model,''
  Phys.\ Rept.\  {\bf 117} (1985) 75.
  %%CITATION = PRPLC,117,75;%%
  %4141 citations counted in INSPIRE as of 29 Apr 2014 
  
 %\cite{Ibrahim:1997gj}
\bibitem{Ibrahim:1997gj}
  T.~Ibrahim and P.~Nath,
  %``The Neutron and the electron electric dipole moment in N=1 supergravity unification,''
  Phys.\ Rev.\ D {\bf 57} (1998) 478
   [Erratum-ibid.\ D {\bf 58} (1998) 019901]
   [Erratum-ibid.\ D {\bf 60} (1999) 079903]
   [Erratum-ibid.\ D {\bf 60} (1999) 119901]
  [hep-ph/9708456].
  %%CITATION = HEP-PH/9708456;%%
  %342 citations counted in INSPIRE as of 30 Apr 2014
    
  %\cite{Falk:1999tm}
\bibitem{Falk:1999tm}
  T.~Falk, K.~A.~Olive, M.~Pospelov and R.~Roiban,
  %``MSSM predictions for the electric dipole moment of the Hg-199 atom,''
  Nucl.\ Phys.\ B {\bf 560} (1999) 3
  [hep-ph/9904393].
  %%CITATION = HEP-PH/9904393;%%
  %172 citations counted in INSPIRE as of 29 Apr 2014
  
  %\cite{Garisto:1996dj}
\bibitem{Garisto:1996dj}
  R.~Garisto and J.~D.~Wells,
  %``Constraints on supersymmetric soft phases from renormalization group relations,''
  Phys.\ Rev.\ D {\bf 55} (1997) 1611
  [hep-ph/9609511].
  %%CITATION = HEP-PH/9609511;%%
  %72 citations counted in INSPIRE as of 29 Apr 2014
  
  %\cite{Lee:2003nta}
\bibitem{Lee:2003nta}
  J.~S.~Lee, A.~Pilaftsis, M.~S.~Carena, S.~Y.~Choi, M.~Drees, J.~R.~Ellis and C.~E.~M.~Wagner,
  %``CPsuperH: A Computational tool for Higgs phenomenology in the minimal supersymmetric standard model with explicit CP violation,''
  Comput.\ Phys.\ Commun.\  {\bf 156} (2004) 283
  [hep-ph/0307377].
  %%CITATION = HEP-PH/0307377;%%
  %231 citations counted in INSPIRE as of 29 Apr 2014
  
    %\cite{Chang:1998uc}
\bibitem{Chang:1998uc}
  D.~Chang, W.~-Y.~Keung and A.~Pilaftsis,
  %``New two loop contribution to electric dipole moment in supersymmetric theories,''
  Phys.\ Rev.\ Lett.\  {\bf 82} (1999) 900
   [Erratum-ibid.\  {\bf 83} (1999) 3972]
  [hep-ph/9811202].
  
 %\cite{Demir:2003js}
\bibitem{Demir:2003js}
  D.~A.~Demir, O.~Lebedev, K.~A.~Olive, M.~Pospelov and A.~Ritz,
  %``Electric dipole moments in the MSSM at large tan beta,''
  Nucl.\ Phys.\ B {\bf 680} (2004) 339
  [hep-ph/0311314].
  %%CITATION = HEP-PH/0311314;%%
  %104 citations counted in INSPIRE as of 29 Apr 2014 
  
  %\cite{Olive:2005ru}
\bibitem{Olive:2005ru}
  K.~A.~Olive, M.~Pospelov, A.~Ritz and Y.~Santoso,
  %``CP-odd phase correlations and electric dipole moments,''
  Phys.\ Rev.\ D {\bf 72} (2005) 075001
  [hep-ph/0506106].
  %%CITATION = HEP-PH/0506106;%%
  %67 citations counted in INSPIRE as of 29 Apr 2014
  
 %\cite{Ellis:2010xm}
\bibitem{Ellis:2010xm}
  J.~Ellis, J.~S.~Lee and A.~Pilaftsis,
  %``A Geometric Approach to CP Violation: Applications to the MCPMFV SUSY Model,''
  JHEP {\bf 1010} (2010) 049
  [arXiv:1006.3087 [hep-ph]].
  %%CITATION = ARXIV:1006.3087;%%
  %11 citations counted in INSPIRE as of 29 Apr 2014 
  
    %\cite{Abel:2001vy}
\bibitem{Abel:2001vy}
  S.~Abel, S.~Khalil and O.~Lebedev,
  %``EDM constraints in supersymmetric theories,''
  Nucl.\ Phys.\ B {\bf 606} (2001) 151
  [hep-ph/0103320].

%\cite{Ellis:2008zy}
\bibitem{Ellis:2008zy}
  J.~R.~Ellis, J.~S.~Lee and A.~Pilaftsis,
  %``Electric Dipole Moments in the MSSM Reloaded,''
  JHEP {\bf 0810} (2008) 049
  [arXiv:0808.1819 [hep-ph]].
  %%CITATION = ARXIV:0808.1819;%%
  %88 citations counted in INSPIRE as of 29 Apr 2014

  %\cite{Kane:2009kv}
\bibitem{Kane:2009kv}
  G.~Kane, P.~Kumar and J.~Shao,
  %``CP-violating Phases in M-theory and Implications for EDMs,''
  Phys.\ Rev.\ D {\bf 82} (2010) 055005
  [arXiv:0905.2986 [hep-ph]].

%\cite{Affleck:1984fy}
\bibitem{Affleck:1984fy}
  I.~Affleck and M.~Dine,
  %``A New Mechanism for Baryogenesis,''
  Nucl.\ Phys.\ B {\bf 249} (1985) 361.
  %%CITATION = NUPHA,B249,361;%%
  %844 citations counted in INSPIRE as of 07 Apr 2014
 

%\cite{Kane:2011ih}
\bibitem{Kane:2011ih}
  G.~Kane, J.~Shao, S.~Watson and H.~-B.~Yu,
  %``The Baryon-Dark Matter Ratio Via Moduli Decay After Affleck-Dine Baryogenesis,''
  JCAP {\bf 1111} (2011) 012
  [arXiv:1108.5178 [hep-ph]].
  %%CITATION = ARXIV:1108.5178;%%
  %12 citations counted in INSPIRE as of 07 Apr 2014

\bibitem{Acharya:2006}
B.~S.~Acharya, K.~Bobkov, G.~L.~Kane, P.~Kumar, and D.~Vaman, Phys.Rev. Lett. \textbf{97}, 191601 (2006), [hep-th/0606262].

\bibitem{Acharya:2007}
B.~S.~Acharya, K.~Bobkov, G.~L.~Kane, P.~Kumar, and J.~Shao, Phys. Rev. \textbf{D76}, 126010 (2007), [hep-th/0701034].

\bibitem{Acharya:2008}
B.~S.~Acharya and K.~Bobkov (2008), [arXiv:0810.3285]

\bibitem{Acharya:2004}
B.~S.~Acharya and S.~Gukov, Phys. Rept. \textbf{392}, 121 (2004), [hep-th/0409191]

  %\cite{Acharya:2010zx}
\bibitem{Acharya:2010zx}
  B.~S.~Acharya, K.~Bobkov and P.~Kumar,
  %``An M Theory Solution to the Strong CP Problem and Constraints on the Axiverse,''
  JHEP {\bf 1011} (2010) 105
  [arXiv:1004.5138 [hep-th]].
  %%CITATION = ARXIV:1004.5138;%%
  %51 citations counted in INSPIRE as of 27 May 2014


%\cite{Witten:2001bf}
\bibitem{Witten:2001bf}
  E.~Witten,
 `` \textit{Deconstruction, G(2) holonomy, and doublet triplet splitting},''
  hep-ph/0201018.
  %%CITATION = HEP-PH/0201018;%%
  %120 citations counted in INSPIRE as of 20 Feb 2014
  
  %\cite{Acharya:2011te}
\bibitem{Acharya:2011te}
  B.~S.~Acharya, G.~Kane, E.~Kuflik and R.~Lu,
  %``Theory and Phenomenology of $\mu$ in M theory,''
  JHEP {\bf 1105} (2011) 033
  [arXiv:1102.0556 [hep-ph]].
  %%CITATION = ARXIV:1102.0556;%%
  %11 citations counted in INSPIRE as of 20 Feb 2014

%\cite{Giudice:1988yz}
\bibitem{Giudice:1988yz}
  G.~F.~Giudice and A.~Masiero,
  %``A Natural Solution to the mu Problem in Supergravity Theories,''
  Phys.\ Lett.\ B {\bf 206} (1988) 480.
  %%CITATION = PHLTA,B206,480;%%
  %722 citations counted in INSPIRE as of 19 May 2014

%\cite{oai:arXiv.org:0811.2417}
\bibitem{oai:arXiv.org:0811.2417}
  J.~J.~Heckman and C.~Vafa,
  %``Flavor Hierarchy From F-theory,''
  Nucl.\ Phys.\ B {\bf 837} (2010) 137
  [arXiv:0811.2417 [hep-th]].
  %%CITATION = ARXIV:0811.2417;%%
  %103 citations counted in INSPIRE as of 10 Feb 2014


%\cite{Ramond:1993kv}
\bibitem{Ramond:1993kv}
  P.~Ramond, R.~G.~Roberts and G.~G.~Ross,
  %``Stitching the Yukawa quilt,''
  Nucl.\ Phys.\ B {\bf 406} (1993) 19
  [hep-ph/9303320].

%\cite{Martin:1993zk}
\bibitem{Martin:1993zk}
  S.~P.~Martin and M.~T.~Vaughn,
  %``Two loop renormalization group equations for soft supersymmetry breaking couplings,''
  Phys.\ Rev.\ D {\bf 50} (1994) 2282
   [Erratum-ibid.\ D {\bf 78} (2008) 039903]
  [hep-ph/9311340].
  %%CITATION = HEP-PH/9311340;%%
  %598 citations counted in INSPIRE as of 21 Apr 2014

%\cite{Weinberg:1989dx}
\bibitem{Weinberg:1989dx}
  S.~Weinberg,
  %``Larger Higgs Exchange Terms in the Neutron Electric Dipole Moment,''
  Phys.\ Rev.\ Lett.\  {\bf 63} (1989) 2333.
  %%CITATION = PRLTA,63,2333;%%
  %339 citations counted in INSPIRE as of 19 May 2014

%\cite{Baron:2013eja}
\bibitem{Baron:2013eja}
  J.~Baron {\it et al.}  [ACME Collaboration],
  %``Order of Magnitude Smaller Limit on the Electric Dipole Moment of the Electron,''
  Science {\bf 343} (2014) 6168,  269
  [arXiv:1310.7534 [physics.atom-ph]].
  %%CITATION = ARXIV:1310.7534;%%
  %31 citations counted in INSPIRE as of 07 Apr 2014

%\cite{Baker:2006ts}
\bibitem{Baker:2006ts}
  C.~A.~Baker, D.~D.~Doyle, P.~Geltenbort, K.~Green, M.~G.~D.~van der Grinten, P.~G.~Harris, P.~Iaydjiev and S.~N.~Ivanov {\it et al.},
  %``An Improved experimental limit on the electric dipole moment of the neutron,''
  Phys.\ Rev.\ Lett.\  {\bf 97} (2006) 131801
  [hep-ex/0602020].
  %%CITATION = HEP-EX/0602020;%%
  %490 citations counted in INSPIRE as of 07 Apr 2014

  \bibitem{Griffith:2009zz}
  W.~C.~Griffith, M.~D.~Swallows, T.~H.~Loftus, M.~V.~Romalis, B.~R.~Heckel and E.~N.~Fortson,
  %``Improved Limit on the Permanent Electric Dipole Moment of Hg-199,''
  Phys.\ Rev.\ Lett.\  {\bf 102} (2009) 101601.
  %%CITATION = PRLTA,102,101601;%%
  %148 citations counted in INSPIRE as of 07 Apr 2014

%\cite{Ellis:1978hq}
\bibitem{Ellis:1978hq} 
  J.~R.~Ellis and M.~K.~Gaillard,
  %``Strong and Weak CP Violation,''
  Nucl.\ Phys.\ B {\bf 150}, 141 (1979).
  
   %\cite{Khriplovich:1985jr}
\bibitem{Khriplovich:1985jr}
  I.~B.~Khriplovich,
  %``Quark Electric Dipole Moment and Induced $\theta$ Term in the {Kobayashi-Maskawa} Model,''
  Phys.\ Lett.\ B {\bf 173} (1986) 193
   [Sov.\ J.\ Nucl.\ Phys.\  {\bf 44} (1986) 659]
   [Yad.\ Fiz.\  {\bf 44} (1986) 1019]. 
  
  %\cite{Ellis:1982tk}
\bibitem{Ellis:1982tk}
  J.~R.~Ellis, S.~Ferrara and D.~V.~Nanopoulos,
  %``CP Violation and Supersymmetry,''
  Phys.\ Lett.\ B {\bf 114} (1982) 231.
  %%CITATION = PHLTA,B114,231;%%
  %362 citations counted in INSPIRE as of 01 Apr 2014
  
 %\cite{Bobkov:2010rf}
\bibitem{Bobkov:2010rf}
  K.~Bobkov, V.~Braun, P.~Kumar and S.~Raby,
  %``Stabilizing All Kahler Moduli in Type IIB Orientifolds,''
  JHEP {\bf 1012} (2010) 056
  [arXiv:1003.1982 [hep-th]].
  %%CITATION = ARXIV:1003.1982;%%
  %34 citations counted in INSPIRE as of 27 May 2014
 																	
  %\cite{Svrcek:2006yi}
\bibitem{Svrcek:2006yi}
  P.~Svrcek and E.~Witten,
  %``Axions In String Theory,''
  JHEP {\bf 0606} (2006) 051
  [hep-th/0605206].
  %%CITATION = HEP-TH/0605206;%%
  %257 citations counted in INSPIRE as of 27 May 2014

%\cite{BowserChao:1997bb}
\bibitem{BowserChao:1997bb}
  D.~Bowser-Chao, D.~Chang and W.~-Y.~Keung,
  %``Electron electric dipole moment from CP violation in the charged Higgs sector,''
  Phys.\ Rev.\ Lett.\  {\bf 79} (1997) 1988
  [hep-ph/9703435].
  %%CITATION = HEP-PH/9703435;%%
  %20 citations counted in INSPIRE as of 20 May 2014
  
  %\cite{Pilaftsis:1999td}
\bibitem{Pilaftsis:1999td}
  A.~Pilaftsis,
  %``Higgs boson two loop contributions to electric dipole moments in the MSSM,''
  Phys.\ Lett.\ B {\bf 471} (1999) 174
  [hep-ph/9909485].
  %%CITATION = HEP-PH/9909485;%%
  %121 citations counted in INSPIRE as of 20 May 2014
  
  %\cite{Chang:1999zw}
\bibitem{Chang:1999zw}
  D.~Chang, W.~-F.~Chang and W.~-Y.~Keung,
  %``Additional two loop contributions to electric dipole moments in supersymmetric theories,''
  Phys.\ Lett.\ B {\bf 478} (2000) 239
  [hep-ph/9910465].
  %%CITATION = HEP-PH/9910465;%%
  %80 citations counted in INSPIRE as of 20 May 2014
  
  %\cite{Pilaftsis:2002fe}
\bibitem{Pilaftsis:2002fe}
  A.~Pilaftsis,
  %``Higgs mediated electric dipole moments in the MSSM: An application to baryogenesis and Higgs searches,''
  Nucl.\ Phys.\ B {\bf 644} (2002) 263
  [hep-ph/0207277].
  %%CITATION = HEP-PH/0207277;%%
  %130 citations counted in INSPIRE as of 20 May 2014
  
  %\cite{Chang:2002ex}
\bibitem{Chang:2002ex}
  D.~Chang, W.~-F.~Chang and W.~-Y.~Keung,
  %``New constraint from electric dipole moments on chargino baryogenesis in MSSM,''
  Phys.\ Rev.\ D {\bf 66} (2002) 116008
  [hep-ph/0205084].
  %%CITATION = HEP-PH/0205084;%%
  %56 citations counted in INSPIRE as of 20 May 2014
  
  %\cite{Chang:2005ac}
\bibitem{Chang:2005ac}
  D.~Chang, W.~-F.~Chang and W.~-Y.~Keung,
  %``Electric dipole moment in the split supersymmetry models,''
  Phys.\ Rev.\ D {\bf 71} (2005) 076006
  [hep-ph/0503055].
  %%CITATION = HEP-PH/0503055;%%
  %51 citations counted in INSPIRE as of 20 May 2014
  
  %\cite{Li:2008kz}
\bibitem{Li:2008kz}
  Y.~Li, S.~Profumo and M.~Ramsey-Musolf,
  %``Higgs-Higgsino-Gaugino Induced Two Loop Electric Dipole Moments,''
  Phys.\ Rev.\ D {\bf 78} (2008) 075009
  [arXiv:0806.2693 [hep-ph]].
  %%CITATION = ARXIV:0806.2693;%%
  %38 citations counted in INSPIRE as of 20 May 2014

 %\cite{ArkaniHamed:2004yi}
\bibitem{ArkaniHamed:2004yi}
  N.~Arkani-Hamed, S.~Dimopoulos, G.~F.~Giudice and A.~Romanino,
  %``Aspects of split supersymmetry,''
  Nucl.\ Phys.\ B {\bf 709} (2005) 3
  [hep-ph/0409232].
  %%CITATION = HEP-PH/0409232;%%
  %399 citations counted in INSPIRE as of 10 Apr 2014

%\cite{Giudice:2005rz}
\bibitem{Giudice:2005rz} 
  G.~F.~Giudice and A.~Romanino,
  %``Electric dipole moments in split supersymmetry,''
  Phys.\ Lett.\ B {\bf 634}, 307 (2006)
  [hep-ph/0510197].
  %%CITATION = HEP-PH/0510197;%%
  %45 citations counted in INSPIRE as of 10 Apr 2014
    
 %\cite{Pospelov:2013sca}
\bibitem{Pospelov:2013sca}
  M.~Pospelov and A.~Ritz,
  %``CKM benchmarks for electron EDM experiments,''
  Phys.\ Rev.\ D {\bf 89} (2014) 056006
  [arXiv:1311.5537 [hep-ph]].
  %%CITATION = ARXIV:1311.5537;%%
  %1 citations counted in INSPIRE as of 08 Apr 2014   
    
 
 

  


\end{thebibliography}
\end{document}